\documentclass[twocolumn,showpacs,preprintnumbers,amsmath,amssymb]{revtex4-1}

\usepackage{graphicx}
\usepackage{dcolumn}
\usepackage{bm}
\usepackage{color}
\usepackage{amsmath}
\usepackage{amsfonts}
\usepackage{natbib}
\usepackage{psfrag}
\usepackage{fancyhdr}
\usepackage{float}

\usepackage{revsymb4-1}

\providecommand\bnabla{\boldsymbol{\nabla}}

\begin{document}




\title{A shallow water model for MHD flows with turbulent Hartmann layers}
\author{Alban Poth\'erat and Jean-Philippe Schweitzer}
\affiliation{
Coventry University, Applied Mathematics Research Centre, Priory Street Coventry CV1 5FB, United Kingdom 
}
\email{alban.potherat@coventry.ac.uk}
\date{4 April, 2011}


\begin{abstract}
We establish a shallow water model for flows of electrically conducting fluids 
in homogeneous static magnetic fields that are confined between two parallel 
planes where turbulent Hartmann layers are present. This is achieved by 
modelling the wall shear stress in these layers using the Prandtl's mixing length model, as did the authors of \cite{albouss00}. The idea for this new model 
arose from the failure of previous shallow water models that assumed a laminar 
Hartmann layer to recover the correct amount of dissipation found in some 
regimes of the MATUR experiment. This experiment, conducted by  the authors of 
\cite{messadek02_jfm}, consisted of a thin layer of mercury electrically driven in differential rotation in a transverse magnetic field. Numerical Simulations 
of our new model in the 
configuration of this experiment allowed us to recover experimental values of 
both the global angular momentum and the local velocity up to a few percent 
 when the Hartmann layer was in a sufficiently well developed turbulent state. 
We thus provide an evidence that the unexplained level of dissipation observed 
in MATUR in these specific regimes was caused by turbulence in the Hartmann 
layers. A parametric analysis of the flow, made possible by the simplicity of 
our model, also revealed that turbulent friction in the Hartmann layer 
prevented quasi-2D turbulence from becoming more intense and limited the 
size of the large scales. 
\end{abstract}

\maketitle

\section{Introduction}
\label{sec:intro}
Geophysical and astrophysical flows such as planetary atmospheres, oceans and 
accretion disks are, to a large extent, governed by 2D dynamics, 
and have been providing a continuous incentive to study 
2D flows for decades. From the theoretical point of view, 2D flows
in general and 2D turbulence in particular offer a realistic and more 
accessible alternative to their 3D counterpart, 
both in terms of complexity and computational costs. Achieving flows with  
purely 2D dynamics in an experiment, however, presents 
somewhat of a challenge, because in the laboratory, nothing is ever quite 2D.
A promising solution emerged when quasi-2D flows were reproduced in small 
scale laboratory experiments (10-20 cm) by imposing a strong enough 
homogeneous static magnetic field ( $B\gtrsim0.1$ T ) across a thin layer of liquid metal (thickness $H\sim1$ cm). The layer was either confined between 
two walls, or between a wall and a free surface, as in Lehnert's experiment 
\cite{lehnert55_prsa} which was probably the first of this kind. In this class 
of experiments, the flow was never strong enough to affect the externally 
imposed magnetic field \cite{roberts67}: the main electromagnetic effect was 
that of the Lorentz force, which diffused the momentum along the field lines. 
For a given structure of size $l_\perp$ and velocity $U_L$, diffusion was 
achieved over the entire channel 
width $H$ in typical time $\tau_{2D}=\rho/(\sigma B^2)(H/l_\perp)^2$ 
\cite{sm82} ($\rho$ and $\sigma$ are the fluid density and electrical 
conductivity). Even for moderately intense flows, this time was much shorter 
than the typical structure turnover time $l_\perp/U_L$, so physical quantities 
were indeed invariant across the channel, except in the boundary layers along 
the channel walls, called Hartmann layers.\\ 
Because of them, this class of flow is not strictly 2D but only \emph{quasi-2D}, and requires dedicated models, such as the SM82 model formulated by 
\cite{sm82}. This model was obtained following the shallow-water approach, a popular technique to model geophysical flows \cite{pedlosky87}.
The idea of shallow water models is that 
when physical quantities vary little in one of the directions 
of space (here $\mathbf e_z$), the fluid motion mostly takes place in the other 
two so it is still well represented by averaging the governing equations along 
the short dimension \cite{pedlosky87}. Theory and numerical simulations based 
on the SM82 model could finely reproduce the details of quasi-2D flows observed in experiments, as long as outside of the boundary layers, the 
momentum diffusion along the magnetic field lines acted much faster than 3D 
inertia and viscous friction. The ratios of these effects are respectively 
measured by two non-dimensional numbers: the 'true' interaction parameter 
introduced by \cite{sreenivasan02}, $N_t=N(l_\perp/H)^2$, and the square of the Hartmann number $Ha=BH(\sigma/(\rho \nu))^{1/2}$. Here the interaction parameter
 $N=\sigma B^2 H/(\rho U)$, where $U$ is a typical fluid velocity  is based on 
the same lengthscale as the Hartmann number for convenience. In these notations,
 the ratio of the Lorentz force to 2D inertia in the core of quasi-2D flows 
such as those we are interested in is of the order of $(N/Ha)(l_\perp/H)$, 
whereas in the Hartmann layer, it becomes of the order of 
$N(l_\perp/H)$ \cite{sm82,psm00}.
The MAgnetic TURbulence (MATUR) experiment in Grenoble
\cite{verron87,delannoy99}, was an experiment where these conditions were well 
satisfied. 
Over the years, its successive versions have been providing a wealth of 
reference data that have motivated the development of models for MHD and 
quasi 2D flows: most recently, \cite{smolentsev07_pf} proposed a model for 
 quasi-2D turbulence under high magnetic fields, based on SM82. 
MATUR consisted of a 
thin, cylindrical container filled with mercury placed in a transverse 
magnetic field (figure \ref{fig:matur}), where a circular, turbulent  shear 
layer was generated by electrically driving into rotation the outer region of 
the cylindrical fluid domain \cite{albouss99}. When $Ha>>1$ and $N(l_\perp/H)>>1$, the Hartmann boundary layers that confined the flow were laminar, with a 
simple exponential profile, as assumed in SM82. For $N(l_\perp/H)\simeq1$, 
however, both global 
and local recirculations at the scale of individual vortices appeared. They  
transfered angular momentum to the side layers at the outer edge of the 
container where an extra dissipation took place that SM82 could not account 
for. This was  later corrected in the more refined PSM model \cite{psm00}. 
This new model included inertia in the Hartmann layers, which was responsible 
for these recirculations and was able to accurately reproduce experimental 
results in these regimes \cite{psm05}. One set of measurements remains, however, where both 
SM82 and PSM grossly underestimate the dissipation. In this regime, 
$Ha\in\{132,212\}$ and $N(l_\perp/H)>>1$, so the 3D recirculations described by PSM are 
too weak to produce the missing dissipation. The Reynolds number based on the 
Hartmann layer thickness $R=UH/(2\nu Ha)=Re/(2Ha)$, however, was over the value of 380, for 
which the Hartmann layer in a rectilinear channel flow becomes turbulent 
\cite{moresco04, krasnov04}. In spite of the difference 
between this ideal configuration and MATUR, it is tempting to think that the 
missing dissipation could be found in turbulent Hartmann layers.\\
In this paper, we explore this possibility by building a 2D model based on the 
assumption of a turbulent Hartmann layer. We shall proceed as follows: we 
first recall the general form of 2D MHD models. We then insert the model for 
turbulent Hartmann layers derived by \cite{albouss00} in this general form to obtain our particular model (section \ref{sec:model}). We then turn our attention 
to MATUR where we obtain a first estimate for the global angular momentum out 
of an axisymmetric version of our new model (section \ref{sec:matur}). Finally, 
we implement our model in the code we previously used to simulate the SM82 
and PSM equations \cite{psm05}, and simulate the flow in MATUR in detail 
(section \ref{sec:num}).
%
\section{Model equations}
\label{sec:model}
\subsection{Shallow water models in Low-$Rm$ MHD}
To establish the shallow water equations, we shall consider the generic 
configuration of an MHD channel flow: an electrically conducting fluid (density 
$\rho$, kinematic viscosity $\nu$, electrical conductivity $\sigma$) is 
confined between two horizontal impermeable walls respectively located at 
$z=-H/2$ and $z=H/2$ and the whole fluid domain is subject to an externally 
applied homogeneous magnetic field $B\mathbf e_z$. We shall work under the 
low-$Rm$ approximation ($Rm=\mu\sigma Ul_\perp<<1$) \cite{roberts67}, valid for liquid 
metals flowing at moderate speeds and in moderately large fluid domains such 
as those encountered in many engineering and laboratory situations. Its main 
implication is that, although the electric current induced by the motion of
conducting fluid in the magnetic field (or order $\sigma BU$) cannot be 
neglected as it participates in the Lorentz force, the magnetic field induced 
by this current ($\sim BRm$) is, by contrast, negligible. Consequently, the 
fluid motion is incapable of modifying the externally applied field and 
electromagnetic effects only appear through the Lorentz force in the momentum 
equations. 
Under this assumption, and normalising lengths by $H$, velocities by $U$, time by $H/U$, pressure by $\rho U^2$, shear stress by $(\rho \nu U/H) Ha$ and 
electric current density by $\sigma B U/Ha$,  
the average along $\mathbf e_z$  of the equations that 
express the conservation of momentum and mass can be written in non-dimensional 
form as \cite{psm05}:
%
\begin{eqnarray}
\partial _{t}\mathbf{\bar{u}}_{\bot }+
\mathbf{\bar{u}}_{\bot }.\bnabla_\bot \mathbf{\bar{u}}_{\bot }
+\overline{\left( \mathbf{u}^{\prime }\mathbf{.\nabla }\right)\mathbf{u}^{\prime }}
+\mathbf{\nabla }_\bot\bar{p} = \nonumber \\
\dfrac{N}{Ha^2}\mathbf{\nabla}^2_{\bot }\mathbf{\bar{u}}_{\bot }
+\frac{N}{Ha}\left(\mathbf{\bar{j}}_\bot\times\mathbf e_z\right)
-2\frac{N}{Ha^2}\mathbf{\tau }_{W},
\label{eq:ns2dmod}\\
\nabla\cdot\overline{\mathbf u}=0,
\label{eq:cont2d}
\end{eqnarray}%
where the over-bar denotes z-averaging across the fluid depth ($z=-1/2$ to $z=1/2$),
 $\mathbf u^\prime$ represents the departure from the averaged velocity 
$\mathbf{ \bar u}$, and $\tau_W(x,y)$ is the friction at a single Hartmann 
wall. At this point, the velocity scale $U$ is left unspecified to keep the 
generality of the model, but will be assigned a value in section \ref{sec:matur}
for the particular case of the MATUR experiment. The governing parameters are 
the Hartmann number $Ha=BH(\sigma/(\rho\nu))^{1/2}$ and the 
interaction parameter $N=\sigma B^2 H/(\rho U)$ introduced in section 
\ref{sec:intro}.
Quantities averaged along $z$  are by definition dependent only on
$x$ and $y$. The corresponding Nabla operator $\bnabla_\bot$ is
2D and carries the subscript ()$_{\bot }$. Similarly, the same
subscript on a vector indicates components perpendicular to the magnetic field 
only. $\mathbf{\bar j}_\bot$ can be expressed by averaging the equations governing the continuity of electric current and Ohm's law:
\begin{eqnarray}
\bnabla_\perp.\mathbf{\bar j}_\bot&=&-j_W,\\
\frac{1}{Ha}\mathbf{\bar j}_\perp&=& \mathbf{\bar E}_\perp+\mathbf{\bar u}_\perp\times \mathbf e_z,
\label{eq:ohm}
\end{eqnarray}
where $j_W$ is the current density injected at one or both of the confining 
planes and $\mathbf E$ is a non-dimensional electric field. Taking the curl of 
the Ohm's law and using the incompressibility condition, one sees that 
$\mathbf{\bar j_\bot}$ is irrotational. It follows that there is a potential 
$\psi_0$ for $\mathbf{\bar j}_\bot$ which satisfies Poisson's equation, the 
source term being $j_W$:
\begin{equation}
\begin{array}{cc}
\mathbf{\bar j}_{\bot }=\mathbf{\nabla }_\bot\psi _{0}, & 
\bnabla_{\bot}^2\psi _{0}=-j_{W}.%
\end{array}
\label{eq:current}
\end{equation}
The potential $\psi_0$ is determined from the current source as the solution of  Poisson's equation (\ref{eq:current}), which is unique for given boundary 
conditions for the electric current 
at the lateral boundaries. Then,
using the vector field $\mathbf u_0$ of streamfunction $\psi_0$, the
 Lorentz force in equation (\ref{eq:ns2dmod}) turns out to
 be only determined by the boundary condition on the electric current 
through (\ref{eq:current}) and $\mathbf{\bar j}_\bot \times  \mathbf e_z=\mathbf u_0$.\\
At this point, the equations have been simply averaged, and no assumption has 
been added to the Navier-Stokes equations. To complete the construction of a 
2D model the averaged equations must be closed by the addition of a model for 
the inertial term $\overline{\left( \mathbf{u}^{\prime }\mathbf{.\nabla }\right)\mathbf{u}^{\prime }}$ as well as for the wall friction term $\mathbf{\tau }_{W}$ in (\ref{eq:ns2dmod}).
\subsection{Model for flows with turbulent Hartmann layers}
\label{sec:2dturb}
\begin{figure}
\includegraphics[width=8.5cm]{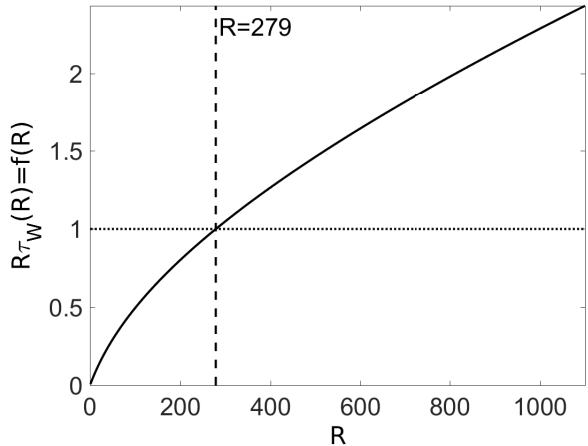}
\caption{Wall friction due to a turbulent (solid line) and laminar 
(dotted line) Hartmann layer, normalised by the latter, \textit{vs.} $R$, the Reynolds number based on the Hartmann layer thickness and core (outer) velocity. This same ratio can also be expressed using dimensional quantities as 
$\tilde\tau_W(U)t_H/(HU)=f(R)$. It is equal to the ratio of laminar to turbulent friction times too.}
\label{fig:tau_nondim}
\end{figure}

To model the MATUR experiment in regimes where the Hartmann layer 
is thought to be turbulent, we shall require two additional assumptions. The 
first one applies to the core of the flow, precisely outside of the Hartmann 
layer (a rigorous definition of this notion can be found in \cite{psm02}). 
There, we shall still assume that the diffusion of momentum along the magnetic 
field lines by the Lorentz force dominates viscous and 3D inertial effects 
outside boundary layers, which is valid in the limit:
\begin{equation}
Ha>>1, \qquad N_t=N\left(\frac{l_\perp}{H}\right)^2>>1.
\label{eq:nondim_cond}
\end{equation} 
The true interaction parameter $N_t$ introduced by \cite{sreenivasan02}, 
represents the effective ratio of the momentum diffusion along magnetic field 
lines due to the Lorentz force, to inertia, as discussed in introduction.
In this limit, the pressure and the velocity components across the magnetic 
field are invariant in the $z$ direction, outside the Hartmann layers.
These assumptions are often referred to as the 2D core flow approximation 
\cite{moreau90}. Recently,
 we have been able to actually observe this flow structure \cite{kpa09_pre}, as
well as the conditions under which two-dimensionality breaks down 
\cite{kp10_prl} in regimes where the Hartmann layer was most likely laminar. 
Yet, in spite of strong theoretical and numerical support \cite{krasnov04} in 
favour of the 
existence of flows where a turbulent Hartmann layers and a 2D core co-exist, 
their experimental evidence is still lacking.\\ 
  In a way, the 2D core approximation justifies the physical relevance of 2D 
models on account that if $\delta$ denotes the 
thickness of the boundary layers along the channel  walls, then the velocity 
outside them, $\mathbf u^c$, a quantity usually measured in experiments 
\cite{kljukin98}, is well approximated by the average velocity as 
$\overline{\mathbf u}= \mathbf u^c+O(\delta/H)=\mathbf u^c+O(Ha^{-1})$. This implies in particular 
that $\overline{\left( \mathbf{u}^{\prime }\mathbf{.\nabla }\right)\mathbf{u}^{\prime }}\sim (\delta/H)^2\mathbf{\bar{u}}_{\bot }.\bnabla_\bot \mathbf{\bar{u}}_{\bot }$. For moderate values of  $N(l_\perp/H)$, this term can account for 
 local secondary flows ignited by the rotation of individual quasi-2D vortical 
structures \cite{psm00}. Here, we shall on the contrary assume that $N(l_\perp/H)$ and $H/\delta\sim Ha$ are large enough to neglect it.\\
We are now only left with the wall friction $\mathbf \tau_W$ to model in order 
to complete our shallow water model. The latter is determined by the structure 
of the Hartmann boundary layer present along the channel walls, the stability 
of which is in turn determined by the Reynolds number scaled on its laminar thickness 
$R=Ha/(2N)=Re/(2Ha)$. In configurations where the bulk velocity is nearly uniform, 
it has been observed both in experiments \cite{moresco04} and numerical 
simulations \cite{krasnov04} that the Hartmann layer was laminar for 
$R\lesssim 380$. In this case, its profile is exponential and $\mathbf \tau_W$ 
takes the form of a linear friction term of dimensional characteristic time 
$t_H=H^2/(\nu Ha)$. The first 2D model for MHD flows, called SM82 after 
\cite{sm82}, essentially relies on this 
assumption. As announced in the introduction, our aim is to model quasi-2D 
flows where the Hartmann layer is turbulent. Although the general behaviour of 
the Hartmann layer may differ from that in idealised configurations with 
uniform bulk velocity, we may infer from this ideal case that the Hartmann 
layer is in a developed turbulent state whenever $R$ significantly exceeds the 
ideal threshold value of $380$.  
Several models exist for the turbulent Hartmann layer: while the early 
approaches of \cite{harris60},  \cite{branover67_mhd} \cite{lykoudis60_rmp} and \cite{lykoudis67_pf} attempted to incorporate the effect of the 
Lorentz force on turbulence within the layer, the authors of \cite{albouss00} 
more recently observed that even when electromagnetic forces were dominant in 
the core ($N>>1$ in our  notations), they were still smaller than inertia 
within the boundary layer when it was turbulent. 
 This enabled them to derive a model for the non-dimensional total stress 
$\tau(z,u^c)$ based on the usual Prandtl mixing-length model 
\cite{schlichting55}. For a given value of the core velocity $u^c$, They showed 
that the non-dimensional stress profile 
$\tau(z,u^c)$ across the Hartmann layer located at $z=z_0$  was solution of an 
ODE, which, using stretched variable $\xi=Ha|z-z_0|$, could be written as:
\begin{eqnarray}
\frac{\partial^2 \tau}{\partial^2 \xi^2}&=&\frac{2.5}{R(u^c)}\frac{\sqrt\tau}{\xi},
\label{eq:tau}\\
\tau\left(\frac{11.3 }{\tau_W(u^c) R(u^c)}\right)&=&\tau_W,
\label{eq:tau_bc1}\\
\lim\limits_{\xi\rightarrow+\infty}\tau(\xi)&=&0,
\label{eq:tau_bc2}
\end{eqnarray}
where $R(u^c)=u^cH/(\nu Ha)$.
The unknown wall stress $\tau_W(u^c)=\tau(z_0,u^c)$ is found by a shooting 
method. Figure \ref{fig:tau_nondim} shows the variations of $\tau_W(u^c)$ 
normalised by the laminar wall stress versus $R$, which is the unique parameter 
this ratio depends on. Ones sees that for $R\simeq1000$, which corresponds 
to the regimes attained in the MATUR experiment, the turbulent Hartmann layer
exerts as much as 2 to 3 times the friction of its laminar counterpart on the flow.
Since Hartmann layer friction is almost the exclusive dissipation mechanism in 
the flow, the total angular momentum can be expected to drop by a similar factor
below \cite{delannoy99}'s prediction, which is based on a laminar Hartmann 
layer.\\

To implement this model for $\tau_W$ in (\ref{eq:ns2dmod}), we shall assume that 
$u^c\simeq\overline u$ on the one hand, and that the validity of the model is 
not affected by the spatial or temporal variations of $\overline u(x,y,t)$ and 
can therefore be applied locally on the other. The numerical solution of 
(\ref{eq:tau}-\ref{eq:tau_bc2}) yields a tabulated function 
$\tau_W=g(\overline u)$ which we shall use directly in (\ref{eq:ns2dmod}). 
Finally, our new 2D model consists of the set of equations:
\begin{eqnarray}
\partial _{t}\mathbf{\bar{u}}_{\bot }+
\mathbf{\bar{u}}_{\bot }.\bnabla_\bot \mathbf{\bar{u}}_{\bot }
+\mathbf{\nabla }_\bot\bar{p} = \nonumber \\
\dfrac{N}{Ha^2}\mathbf{\nabla}^2_{\bot }\mathbf{\bar{u}}_{\bot }
+\frac{N}{Ha}\left(\mathbf u_0
-\frac2{Ha}g(\|\mathbf{\bar{u}}_{\bot}\|)\frac{\mathbf{\bar{u}}_{\bot}}{\|\mathbf{\bar{u}}_{\bot}\|}\right)
\label{eq:ns2d}\\
\nabla\cdot\overline{\mathbf u}=0,
\label{eq:cont2d}
\end{eqnarray}%
where $\mathbf u_0$ is built from the streamfunction $\psi_0$, solution of 
(\ref{eq:current}), which is uniquely determined by the electric boundary 
conditions of the problem.
\subsection{2D model with a threshold for the friction}
The model we just established assumes that the Hartmann 
boundary layers are everywhere turbulent. Although this assumption would seem 
reasonable in high speed duct flows, it is more questionable in flows in 
rotation, as in MATUR, where velocities are very low near the centre of 
rotation. This raises the much wider question of the spatial instability of the 
Hartmann layer: in a domain where regions of high velocity where 
$R(\bar u)=\bar uH/(\nu Ha)>380$ and regions of low velocity where 
$R(\bar u)<380$ coexist, can the Hartmann layer be turbulent in the former and 
laminar in the latter? Do, on the contrary, regions of turbulent Hartmann 
layers contaminate those of low velocity where the layer would otherwise be 
laminar? To our knowledge, these questions have not been studied. They 
certainly exceed the scope of our paper, as does the precise modelling of flows where such regions of high and low velocities coexist. 
Since, however, the state of the Hartmann layers may not always be known 
\emph{a priori} in MATUR, we propose a variant to the "all turbulent" model 
from section \ref{sec:2dturb}, where a threshold $R_T$ on the value of 
the parameter $R(\bar u)$ based on the local velocity separates laminar from 
turbulent values of the friction:
\begin{eqnarray}
\tau_W&=&Ha\mathbf u_\perp \quad  {\rm for} \quad R\leq R_T \nonumber\\
\tau_W&=&\frac{N}{Ha}g^{-1}(\|\mathbf u_\perp\|)\frac{\mathbf u_\perp}{\|\mathbf u_\perp\|}  \quad  {\rm for} \quad R> R_T
\end{eqnarray}
In the forthcoming calculations, we set $R_T$ either to the value of 279, at 
which turbulent friction matches laminar friction or of 380 at which Hartmann 
layers become turbulent in duct flows. $R_T=279$ is also close to 
the value at which turbulent Hartman layers re-laminarise \cite{lingwood99}. 
Clearly, the value 
and the very existence of such a threshold do not take their origin in the 
actual physics of the flow. The main advantage of a model with threshold is 
that it is justified both in the limits of low velocities where the Hartmann 
layers are laminar everywhere and of high velocities where they are turbulent 
nearly everywhere. 
%
\section{The MATUR experiment}
\label{sec:matur}
\subsection{Problem geometry}
\begin{figure}
\psfrag{B}{$\mathbf B$}
\psfrag{er}{$\mathbf e_r$}
\psfrag{ez}{$\mathbf e_z$}
\psfrag{O}{O}
\psfrag{H2}{$H/2$}
\psfrag{ri}{$\tilde r_i$}
\psfrag{r0}{$\tilde r_0$}
\psfrag{Iinj}{$I_{inj}$}
\psfrag{Fl}{Liquid metal}
\includegraphics[width=8.5cm]{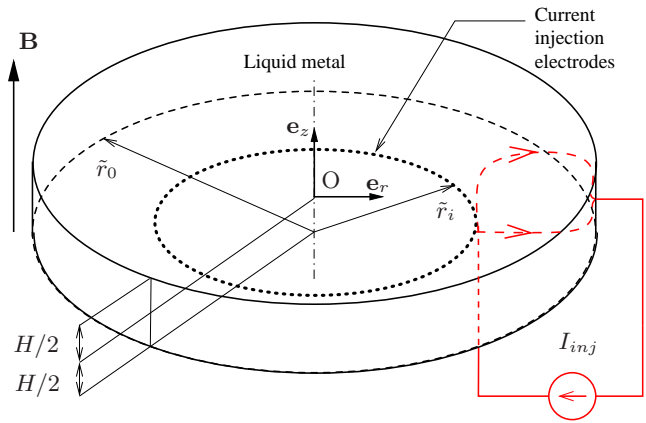}
\caption{Sketch of the Matur Experiment. A typical electric circuit including
one of the point-electrodes mounted flush at the bottom Hartmann layer is 
represented in red. In reality, all electrodes located at $r=r_i$ are 
connected.}
\label{fig:matur}
\end{figure}

We shall now describe the MATUR experiment which inspired the development of our model in the first place. The full detail of the experimental apparatus is 
reported in \cite{messadek02_jfm} and \cite{messadek01_phd}. It consists of an 
airtight cylindrical container of radius $\tilde r_0=11$ cm and depth $H=1$ 
cm entirely filled with mercury ($\rho=1.3529\times10^4$ kg.m$^{-3}$, 
$\nu=1.1257\times10^{-7}$ m$^2$.s$^{-1}$ and 
$\sigma=1.055\times10^6$ $\rm \Omega^{-1}$.m$^{-1}$), and placed in the 
bore of a solenoidal magnet that maintains an homogeneous magnetic field of 
up to 6 T oriented along the cylinder axis $\mathbf e_z$ (the "tilde" indicates 
that quantities are dimensional). The frame origin is 
placed at the centre of the cylinder.  Fluid motion is 
driven by connecting the positive pole of a DC electric current power supply 
to a large number of equally resistive  electrodes mounted flush at the bottom 
wall along a circle of radius $\tilde r_i=5.4$ cm. The negative pole is connected to the
 electrically conducting circular side wall, while Hartmann walls, orthogonal to $\mathbf e_z$ are electrically insulating, except at the locus of the current 
injection electrodes. A simplified sketch of the experiment is shown in figure 
\ref{fig:matur}. Under these conditions, the dimensional injected current 
density at the wall $\tilde j_W$ is axisymmetric and may be modelled to a very 
good approximation as
%
$\tilde j_W=\delta_D(r-r_i)I/(2\pi \tilde r_i)$,
%
where $I$ is the intensity of the total injected current and $\delta_D$ is the 
Delta-Dirac distribution. Solving (\ref{eq:current}) as in \cite{verron87} leads to the expression of 
the dimensional $z-$average of the Lorentz force:
\begin{equation}
\tilde{ \mathbf j_\perp}\times\mathbf B=\rho \frac{\Gamma}{t_H}\mathcal H(\tilde r-\tilde r_i)\frac1{\tilde r},
\mathbf e_\theta
\label{eq:fdim}
\end{equation}
where $\Gamma=I/(2\pi\sqrt{\sigma\rho\nu})$ is the total circulation induced by the current injection, and $\mathcal H(r-r_i)$ is the Heaviside step function. 
The problem geometry and the expression of the forcing suggest the choice of 
$U=\Gamma/\tilde r_0$ as the reference velocity so that the forcing is 
expressed non-dimensionally in (\ref{eq:ns2d}) as:
\begin{equation}
\left(\mathbf{\bar{j}}_\bot\times\mathbf e_z\right)
= \mathbf u_0=\frac{\tilde r_0}{H}\mathcal H(r-r_i)\frac1r\mathbf e_\theta.
\label{eq:u0}
\end{equation}
(\ref{eq:u0}) expresses that the electric current mostly flows radially in the 
Hartmann layers between $r_i$ and $r_0$ so the Lorentz force is azimuthal and 
acts almost exclusively in this region, and not within the disk $r<r_i$ where 
the fluid isn't directly stirred. Initially, the MATUR experiment was indeed  
designed to study the circular shear layer that separates these two regions.
\subsection{An approximate expression for the Angular momentum in MATUR}
\label{sec:axi}

\begin{figure}
\includegraphics[width=8.5cm]{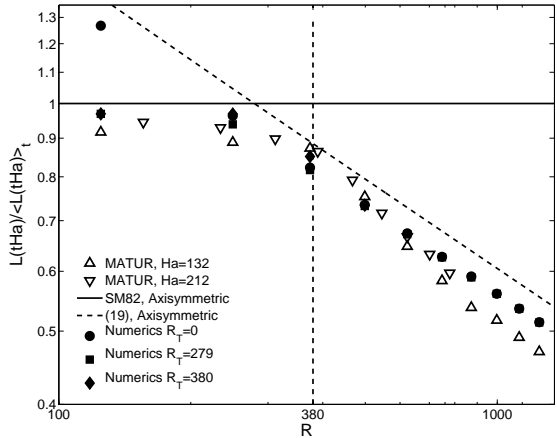}
\caption{Global angular momentum in MATUR for $Ha=132$ and $Ha=212$. 
The axisymmetric approximation is obtained from 
(\ref{eq:laxi}). The critical value for the destabilisation of a plane Hartmann 
layer \cite{moresco04} is marked with a vertical dashed line.}
\label{fig:l}
\end{figure}

Most of the viscous and Joule dissipation in quasi-2D flows under
 strong magnetic field takes place in the Hartmann layers. Whether these layers are laminar or turbulent therefore directly affects the global dissipation. 
In the MATUR experiment, this effect is best revealed through the relation 
between the total injected current and the global angular momentum. As a first 
application of our 2D model, we shall find an approximate relation between 
these two quantities under the simplified assumption that the flow is steady 
and axisymmetric. The total angular momentum can be expressed as:
\begin{eqnarray}
L&=&\int\limits_\Omega r u_\theta(r)d\Omega\\
&=&\int\limits_{0\leq r<r_i} ru_\theta(r)d\Omega+\int\limits_{r_i\leq r \leq r_0} r u_\theta(r)d\Omega.
\end{eqnarray}
Since the most intense part of the flow takes place in the region $r_i\leq r \leq r_0$, 
where the forcing acts, we shall neglect the contribution of the first integral 
to the total angular momentum. Then, by virtue of the mean value theorem, the 
second integral can be related to the azimuthal velocity at a point $r_1$ such 
that $r_i<r_1<r_0$:
\begin{equation}
L=\pi r_1u_\theta(r_1)(r_0^2-r_i^2).
\label{eq:lu}
\end{equation}
The value of $u_\theta(r_1)$ can be estimated using the azimuthal component 
of the Navier-Stokes equation (\ref{eq:ns2d}), by noticing that outside the 
boundary layers, the forcing is mostly balanced by the Hartmann layer friction 
term:
\begin{equation}
u_\theta(r_1)\simeq g^{-1}\left(\frac{Ha}{2r_1}\right).
\end{equation}
%
%
%
%
Since the radial profiles of azimuthal velocity measured in MATUR suggest that 
the local angular momentum $r u_\theta(r)$ 
only slightly increases over $r_i<r<r_0$ (this is confirmed by the radial 
profiles of azimuthal velocity obtained from numerical simulations on figure 
\ref{fig:vtheta} and \ref{fig:vtheta2}), we shall assume that $r_1 u_\theta(r_1)\simeq r_0u_\theta(r_0)$. 
Using (\ref{eq:lu}), 
an estimate 
for the total angular  momentum can be expressed in terms of tabulated function 
$g$ as:
\begin{equation}
L\simeq\pi(r_0^2-r_i^2) g^{-1}\left(\frac{Ha}{2 r_i}\right).
\label{eq:laxi}
\end{equation}
Note that in the case where the Hartmann layers are laminar, the SM82 model 
provides an explicit expression of the angular momentum for axisymmetric 
flows in MATUR as $L_{SM82}=4\pi (r_0^2-r_i^2)$ \cite{delannoy99}.
The values of $L$ obtained under this approximation and (\ref{eq:laxi}) are 
plotted on figure \ref{fig:l}, along with the values of the angular momentum 
measured in 
MATUR for $Ha=132$ and $Ha=212$. We have plotted the original dimensional data 
of \cite{messadek02_jfm} under the form of the angular momentum normalised by 
$L_{SM82}$ 
\emph{vs.} $R$. In these variables, experimental $L(R)$ curves obtained 
at both values of $Ha$ collapse well into a single curve. The 
most important feature of this curve is the rather sharp change of slope around
 $R\simeq 380$. For $R<380$, the experimental values remain reasonably close to 
the SM82 linear approximation.
By contrast, as soon as $R>380$, they fall to significantly lower values than 
the linear prediction. This reveals a much higher level of dissipation in the 
flow than that induced by the
 laminar Hartmann friction, as would be expected when the Hartmann layers become
 turbulent. The value of $R\simeq380$ at which this transition occurs for both 
values of $Ha$ brings support 
to this hypothesis. Even so, it is somewhat remarkable that the transition does 
take place roughly at the same value of $R$ in such strongly different flows as 
 channel flows with only one component of velocity such as the azimuthal flow 
studied by \cite{moresco04} or the rectilinear flow of \cite{krasnov04} on one 
side, and that in MATUR on the other. 
The variations of $L(R)$  calculated with our simplified axisymmetric 
model also support the hypothesis that the Hartmann layers become turbulent in 
MATUR when $R\gtrsim380$, as it reproduces well the trend of the experimental 
values at large $R$: while $L$ is overestimated by 10-20\%, (\ref{eq:laxi}) 
exhibit nearly the same slope as the experimental curve. This level of 
discrepancy is similar to that found in regimes where the Hartmann layer is 
laminar between experimental values and the axisymmetric approximation based on SM82.
Most importantly, for 
$R>380$, where the model is supposed to be valid, (\ref{eq:laxi}) does 
reproduce the extra dissipation, while the linear model doesn't. Based on 
this encouraging result, we shall now lift the limitations of the axisymmetric 
assumption and attempt a more refined description of the flow based on 2D 
numerical simulations of our model.

\section{Numerical simulations of the MATUR experiment}
\label{sec:num}
\subsection{Numerical system and procedure}
The numerical system we use to solve the 2D equations (\ref{eq:ns2d}-\ref{eq:cont2d}) in the MATUR geometry relies on commercial code FLUENT where
the Finite Volumes method is implemented. The code differs very little 
from the one we previously used to simulate flows in the MATUR experiment at 
lower magnetic fields, and the meshes are identical. This earlier work is 
reported in \cite{psm05}, where the code is described in detail and extensively 
tested by following the procedure put forward by \cite{roache97} to measure 
numerical convergence. Further tests on the same solver for the configuration 
of the flow past a cylinder can be found in \cite{dp08}. To briefly 
summarise it, the spatial discretisation is of second order, upwind. The cases 
studied are unsteady and the time-scheme is a second order implicit 
pressure-velocity formulation. Within each iteration,  the equations are solved 
one after the other (segregated mode) using the PISO predictor-corrector 
algorithm proposed by \cite{issa85} to handle the pressure-velocity coupling.
The turbulent Hartmann friction term is treated explicitly at each iteration.
The values of $g(\|\mathbf u\|)$ are interpolated from a table that is 
pre-established by solving (\ref{eq:tau}-\ref{eq:tau_bc2}) for a discrete set of 1100 regularly spaced values of $\|\mathbf u\|$, between 0 and a maximum value of $0.8$.\\
The mesh is made of quadrilateral elements, unstructured for $r<0.15$ and 
structured for $0.15<r<1$. The radial resolution is of 105 points, 25 of which are devoted to the boundary layer located at $r=1$. These points are spread in the layer according to 
 a geometric sequence of ratio 1.3 starting at $r=1$ with an initial interval
of $4.54\times10^{-5}$.  The azimuthal resolution is of 150 points. The time step is
chosen so that the related cutoff frequency matches the spatial cutoff frequency
for the maximum flow velocity (Courant-Friedrich-Lewy condition). The usual no-slip condition at the wall $r=1$ is applied.
\begin{table}
\begin{tabular}{|p{2.5cm}|p{1cm}p{1cm}p{1cm}p{1cm}p{1cm}|}
\hline
$I$ /A          		&10	&20     &30	&40     &50	\\
$\Gamma/(2\tilde r_0)$ /m/s 	&0.182	&0.364  &0.546	&0.728  &0.910\\
$2N$             		&11.71	&5.85   &3.90	&2.93  	&2.34	\\
$R$				&125	&249	&374	&499	&623\\
time step $\times 10^{-4}$
				&2.5	&2.6	&5.3	&4.0	&5.3\\
\hline
$I$ /A          		&60     &70     &80     &90     &100\\
$\Gamma/(2\tilde r_0)$ /m/s    	&1.09   &1.27	&1.45   &1.64	&1.82\\
$2N$             		&1.95   &1.67	&1.46   &1.30	&1.17\\
$R$             		&748	&872    &997    &1122	&1247\\
time step $\times 10^{4}$
				&5.0  	&5.8	&4.6	&5.2	&5.8\\
\hline
\end{tabular}
\caption{Dimensional and non-dimensional parameters for the 2D simulations of 
the MATUR experiment for $Ha=132$. The non-dimensional time step is normalised 
by $\Gamma^{-1}$. It should be noted that the velocity estimate 
$\Gamma/(2\tilde r_0)$ only gives an accurate estimate of the actual flow 
velocity when the Hartmann layer is laminar (this can be seen on figure 
\ref{fig:l}). Turbulent dissipation in the Hartmann layer considerably reduces 
the latter for $R>380$, so that in this regime, $2N$, which is an interaction 
parameter  based on $\Gamma/(2\tilde r_0)$  is noticeably lower than an interaction parameter that 
would be based on true values of the core velocity, and conditions 
(\ref{eq:nondim_cond}) are comfortably satisfied.}
\label{tab:param}
\end{table}
All calculated cases are listed in table \ref{tab:param}, with their  
corresponding non-dimensional parameters and time steps. The flow is initially 
at rest while the forcing is constant, given by (\ref{eq:u0}) for $t\geq0$.\\

Since the velocities involved in the cases simulated in the present work are 
considerably higher than those in \cite{psm05}, the suitability of our mesh 
(which we shall denote M$_1$) was tested by comparing the numerical solution 
obtained with it for $Ha=132$ and $R=1122$ to one obtained with a mesh with the same 
structure, but where 
the resolution was doubled both in the radial and the azimuthal directions 
(mesh M$_2$). The time-averaged global angular momentum and $\mathcal L^2$ norm 
of the error on  azimuthal velocities 
in the established state are gathered in table \ref{table:convergence}. The 
relative discrepancy between the two solutions remains around 1\%
 (see profiles on figure \ref{fig:vtheta2}).
In view of these results, we deem M$_1$ suitable for the 
problem we investigate.
\begin{table}
\begin{tabular}{|p{2cm}|p{2cm}p{3cm}|p{2cm}|}
\hline
   	&$\frac{L}{L^{\rm SM82}}$	
&$\frac{\|<u_\theta>_t-<u_\theta^{(M_2)}>_t\|_2}{\|<u_\theta^{(M_2)}>_t\|_2}$\\
\hline
Mesh M$_1$ 	&0.5355  &0.0198 \\
Mesh M$_2$      &0.5375   &0  \\
\hline
\end{tabular}
\caption{Comparison between simulations performed on meshes M$_1$ and M$_2$ for $Ha=132$ and $R=1122$.}
\label{table:convergence}
\end{table}
%
%
\subsection{General aspect of the flow}
\label{sec:aspect}
\begin{figure}
\begin{center}
\begin{tabular}{cc}
$tHa=0.073$&	$tHa=0.146$\\
\includegraphics[width=4.25cm]{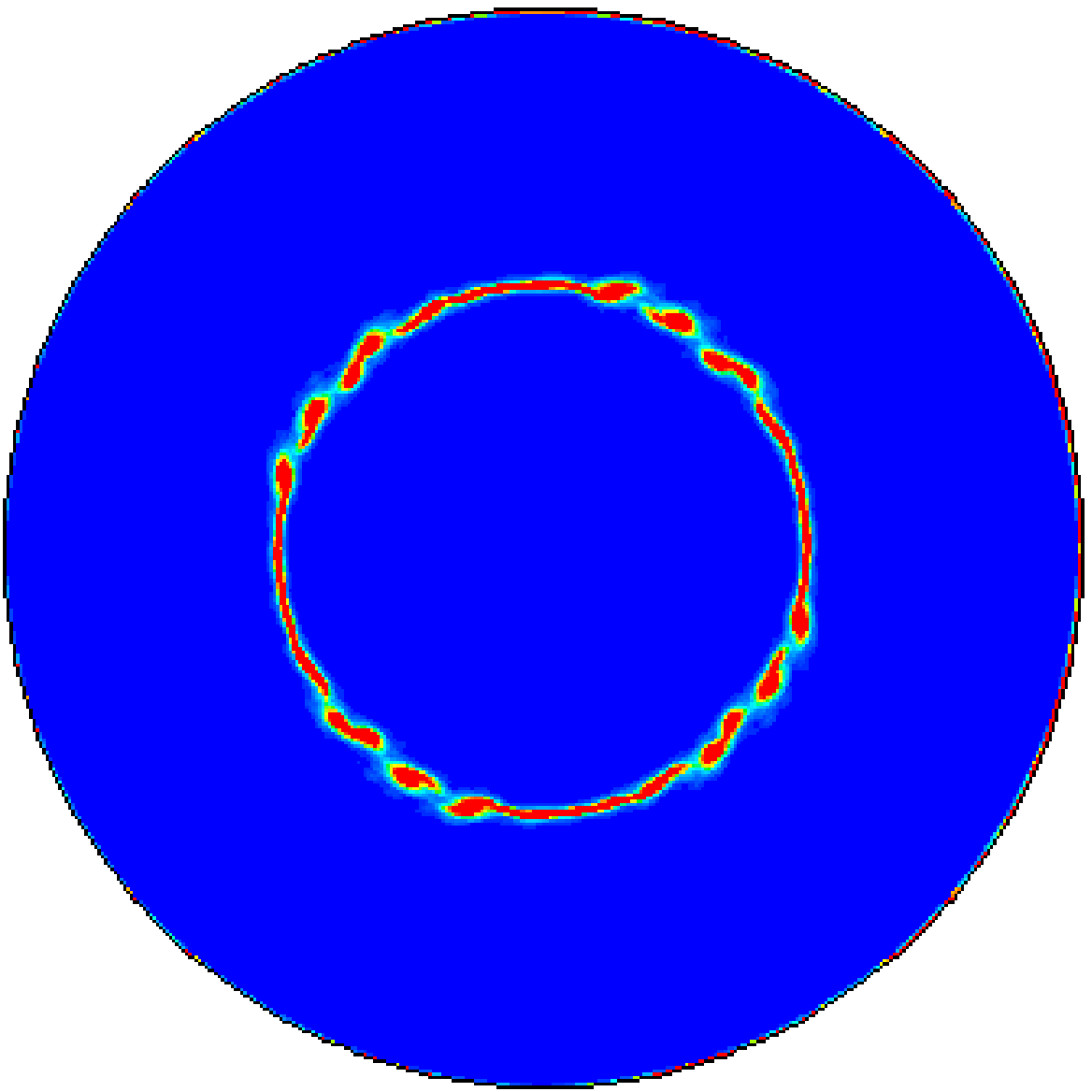}&
\includegraphics[width=4.25cm]{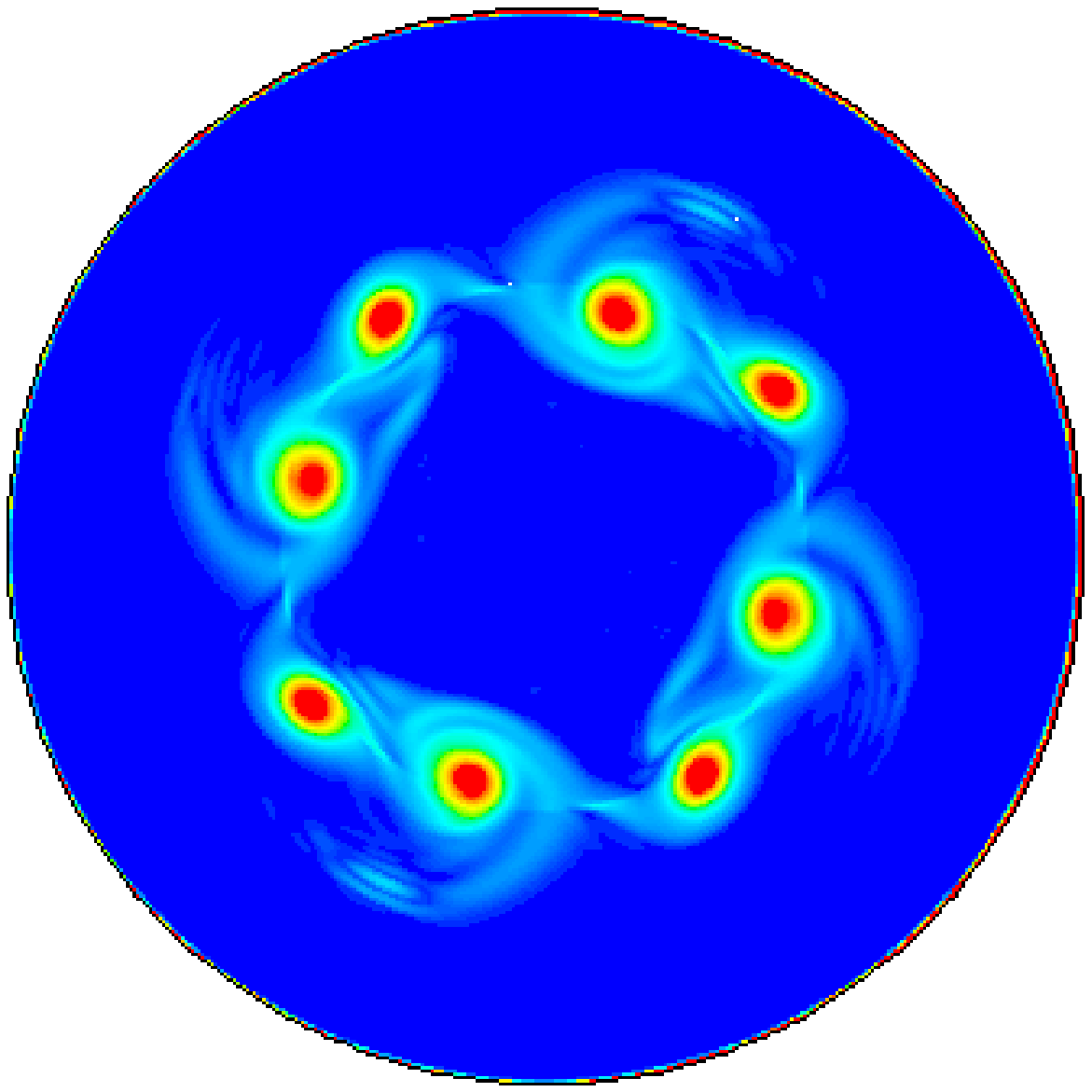}\\
$tHa=0.292$	&	$tHa=0.366$\\
\includegraphics[width=4.25cm]{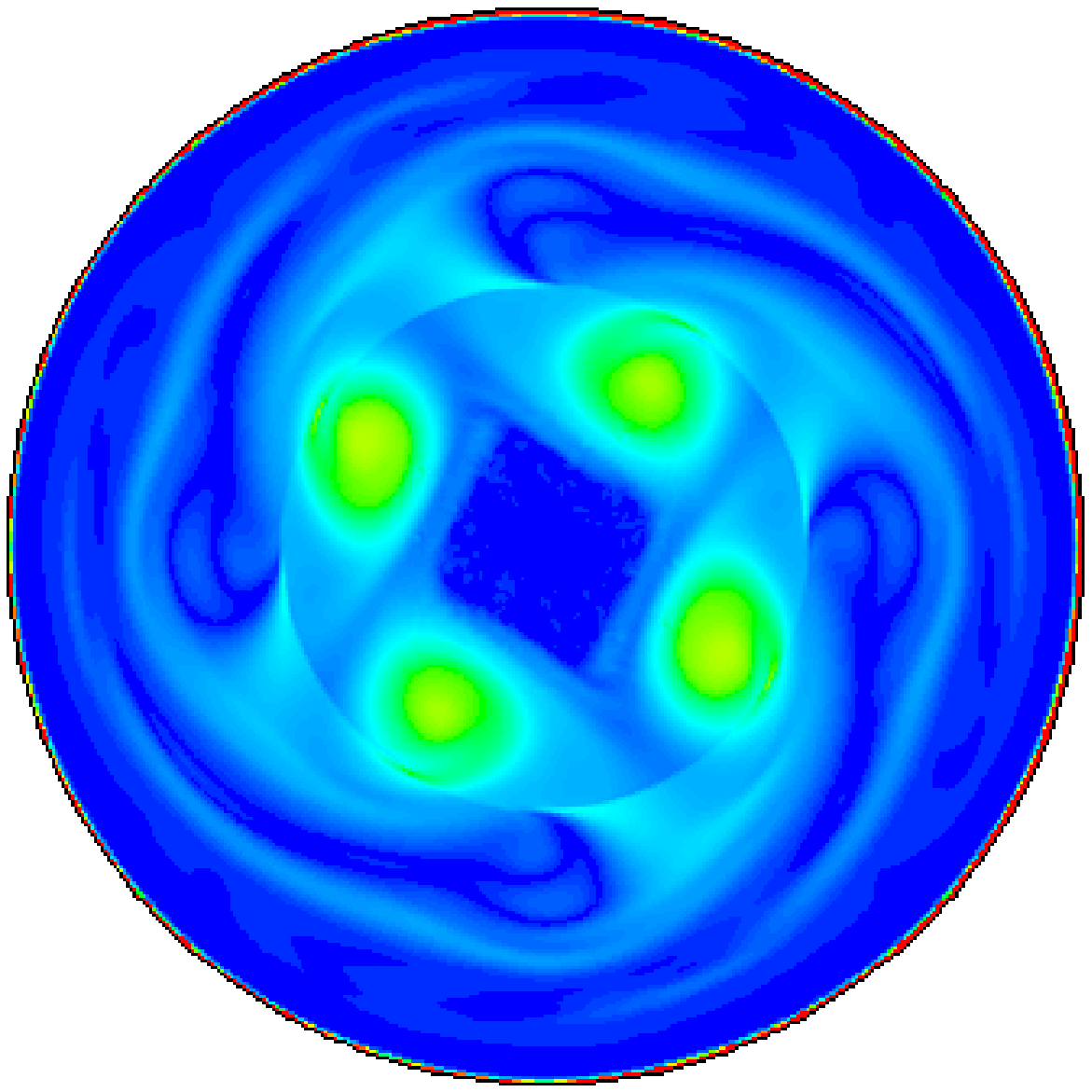}&
\includegraphics[width=4.25cm]{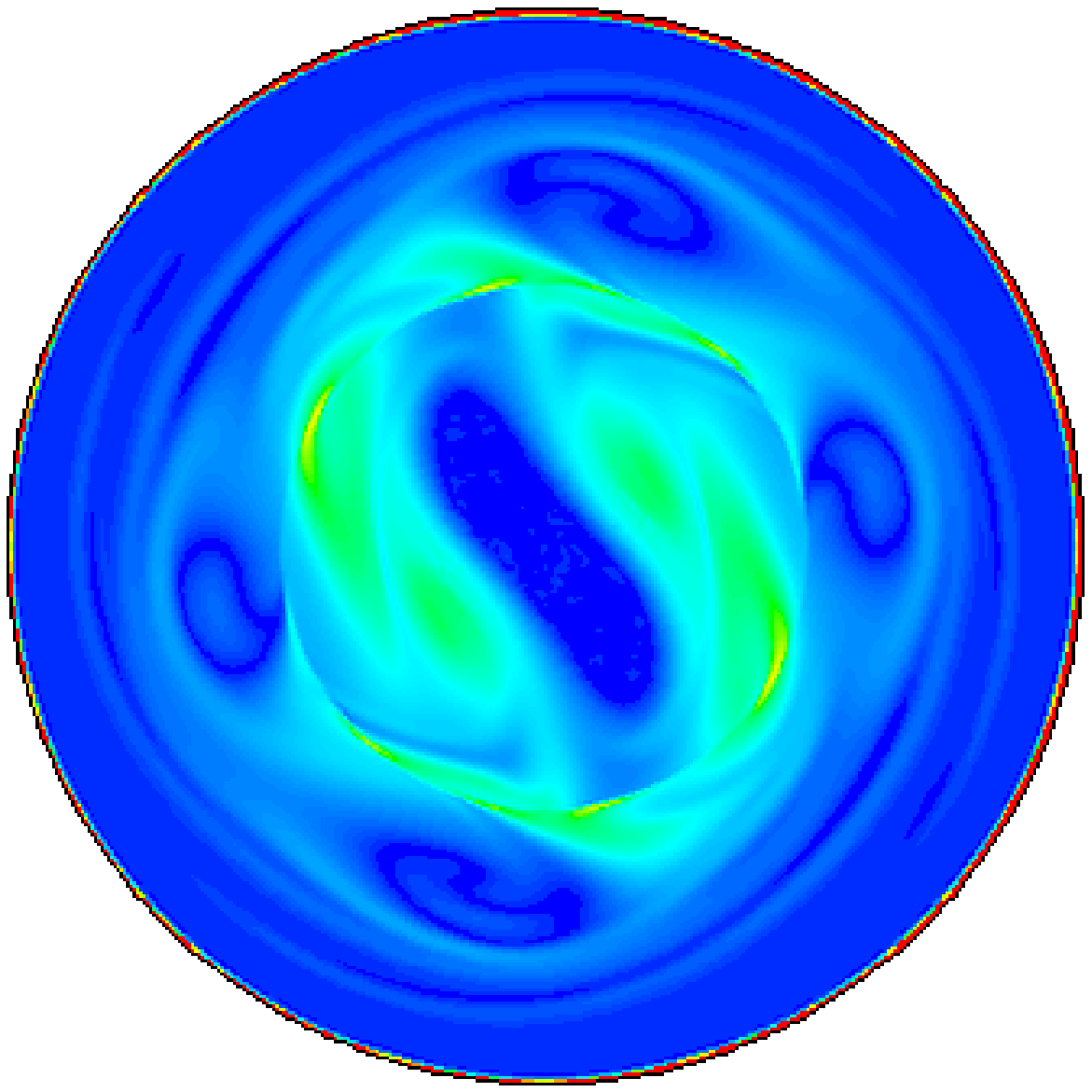}\\
$tHa=0.511$     &       $tHa=2.34$\\
\includegraphics[width=4.25cm]{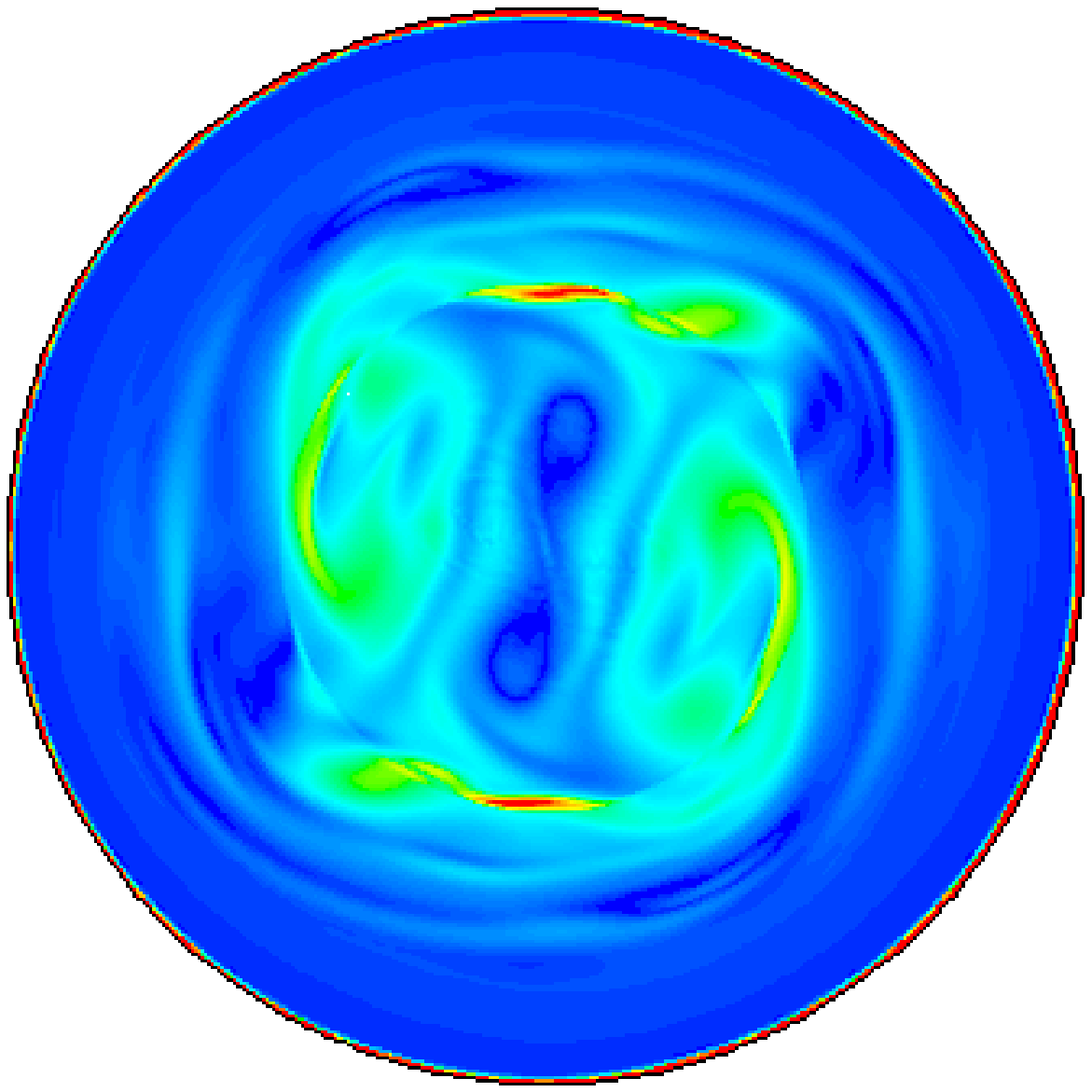}&
\includegraphics[width=4.25cm]{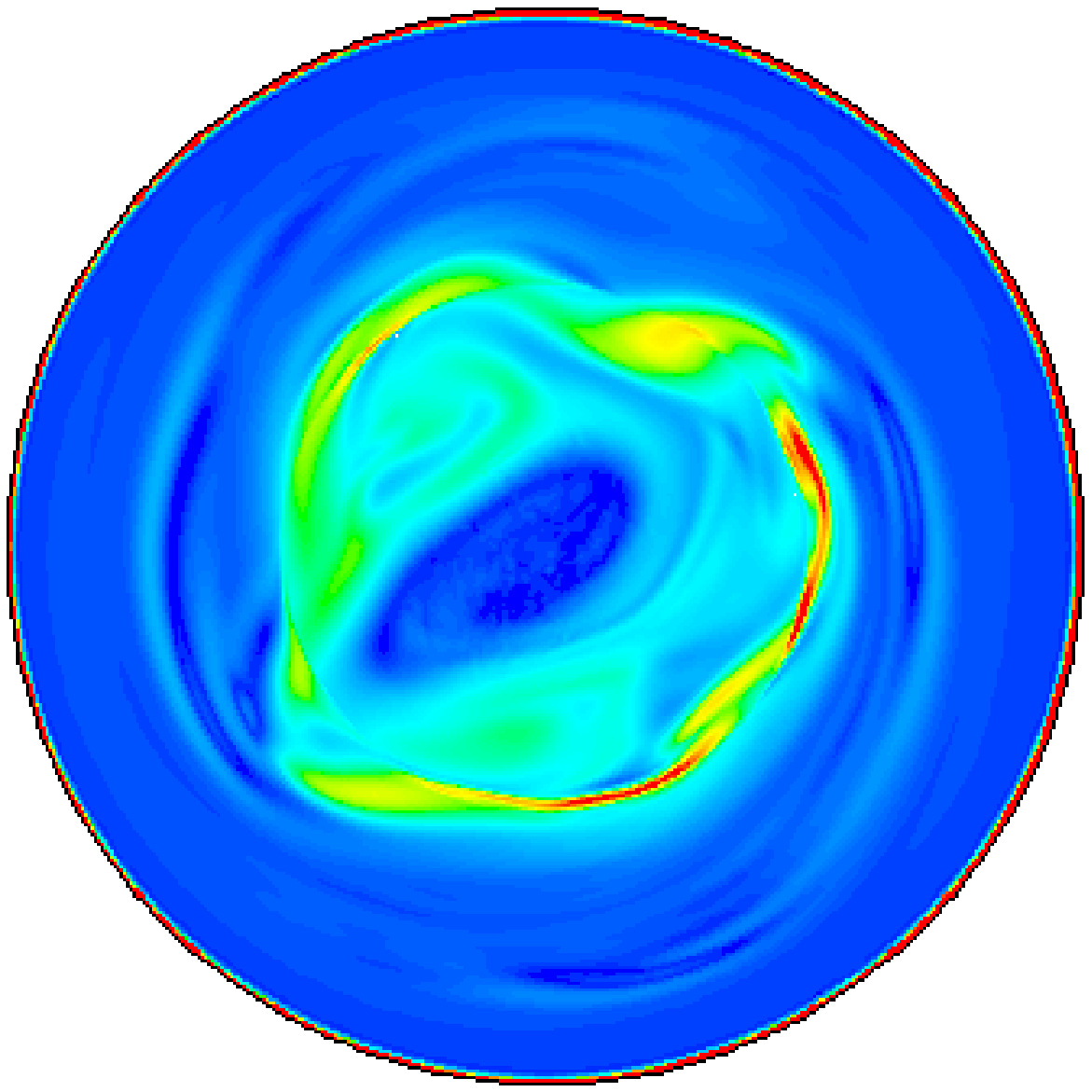}\\
$tHa=2.41$     &       $tHa=2.56$\\
\includegraphics[width=4.25cm]{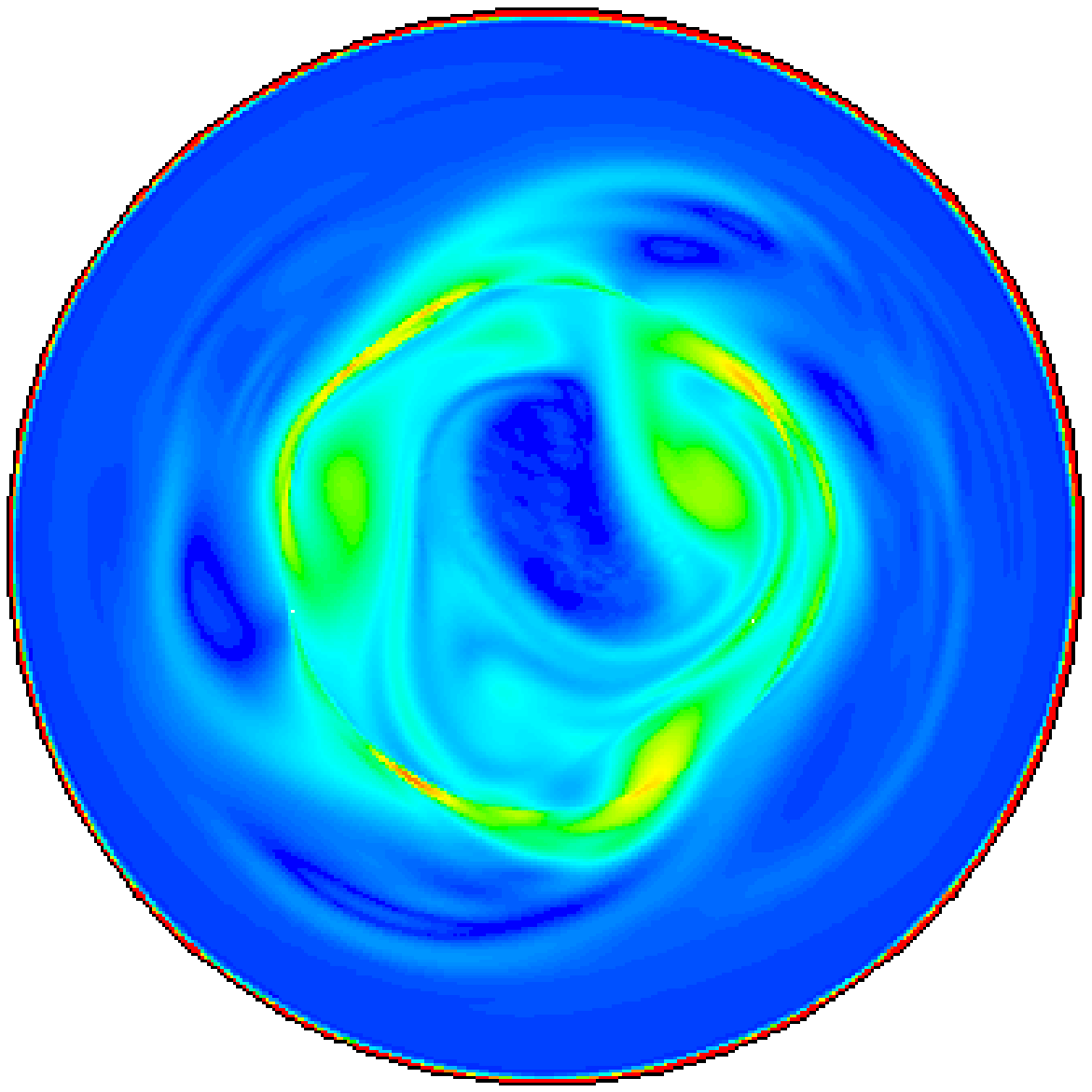}&
\includegraphics[width=4.25cm]{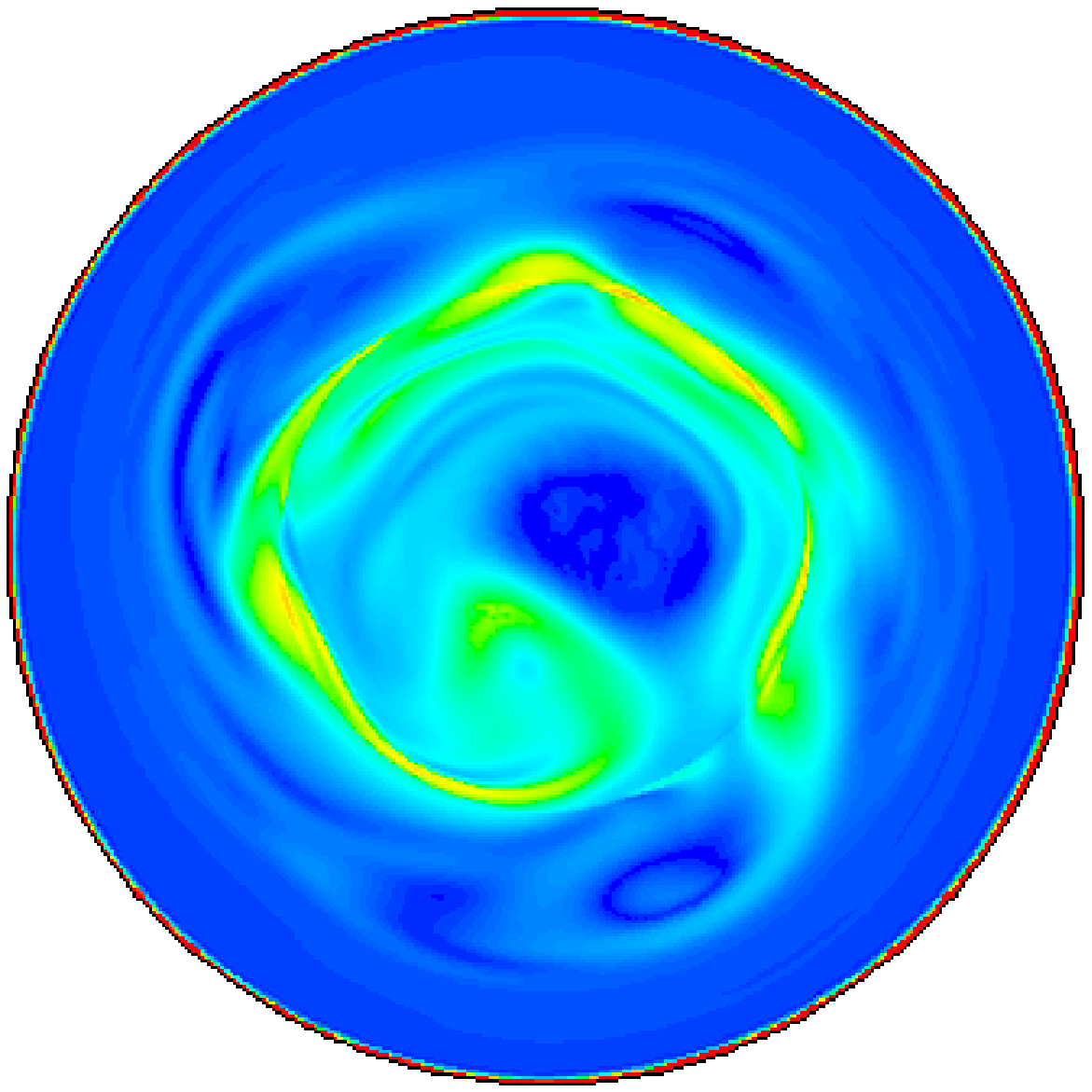}\\
\end{tabular}
\\
\psfrag{0}{\tiny 0}
\psfrag{wm}{$\omega_{\rm max}$}
\psfrag{w2}{$\omega_{\rm max}/2$}
\includegraphics[width=4cm]{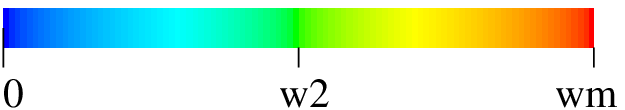}
\caption{Evolution of the flow from rest when the forcing is switched on for 
$Ha=132$ and $R=1122$, obtained from numerical simulations based on the 2D model with $R_T=0$. 
}
\label{fig:w_contours}
\end{center}
\end{figure}

The evolution of the flow is qualitatively similar to that found in our 
previous simulations of MATUR at lower $Ha$, where the current was injected closer to the wall (In \cite{delannoy99} and \cite{psm05}, $r_i/r_0=0.845$  and the Hartmann layer remained laminar.). Its main stages are represented by contours of 
vorticity on figure \ref{fig:w_contours}. At first, a laminar shear layer 
appears at $r=r_i$ 
as the external corona $r_i\leq r <r_0$ is driven in rotation. For all 
intensities of total injected current considered here, a threshold on the 
azimuthal velocity is very quickly reached where this circular free shear is 
subject to a Kelvin-Helmholtz 
instability that breaks it up into small vortices. These soon begin to merge 
into larger structures. They become distorted by the shear and the flow turns 
chaotic before it reaches a final turbulent state. Injecting the electric 
current at a lower radius than in the cases studied in \cite{delannoy99,psm05} 
introduces two differences: 
firstly, most large vortices and associated turbulent fluctuations remain relatively close to the centre of the domain, which unlike when $r_i/r_0=0.845$, is not still, but subject to a highly fluctuating fluid motion. Conversely, velocity 
fluctuations in the region near the outside cylinder wall are of much lower 
intensity. They result mostly from the tail of vortices generated near the 
injection electrodes that are stretched by the shear and conveyed outwards. 
The resulting flow in the outer region therefore exhibits long azimuthal 
vorticity streaks of much lower intensity than in the disk inside the circle of
injection electrodes. Also, since large structures do not reach the outer 
wall,  no flow separation occurs there.  This wall has thus little 
influence on the flow, unlike in the two previously mentioned studies 
where the current was injected closer to it.\\
When the flow is well established, it goes through a recurring sequence.
In the first phase, very strong vorticity emerges in segments along the circle 
where the current is injected (see figure \ref{fig:w_contours}, $tHa=2.34$). In 
the second phase, these fragile segments break up and roll into vortical 
structures ($tHa=2.41$). Those merge in the third phase to build up a small 
number of larger structures (at least two, as at $tHa=2.56$). These large 
structures progressively loose intensity as the cycle returns to the 
first phase.
\subsection{Global angular momentum}
The presence of large vortices carried by the flow has a direct impact on the 
global angular momentum. Figure \ref{fig:l} indeed shows that the time averaged 
angular momentum computed in the established regime from the numerical 
simulations stands a little below the axisymmetric approximation of 
section \ref{sec:axi}, which ignored these large vortices. Remarkably, it 
stands on a curve that is closely parallel to that of the axisymmetric 
approximation but improves it by bringing the discrepancy to experimental 
values below 10\% in the limit of large $R$. This remaining discrepancy may
not even necessarily be attributed to the 2D model as \cite{messadek02_jfm} 
point out that metallic electrodes embedded in one of the rig's Hartmann walls 
precisely incur about 10\% extra dissipation on the flow. Since this extra 
dissipation is not accounted for in either SM82 or PSM, the authors suggest 
that it may explain the discrepancy between experimental values and those 
obtained with SM82 in regimes where the Hartmann layer is laminar. It is thus 
reasonable to expect that the same mechanism is at play when the Hartmann layer 
is turbulent.\\
It is not surprising 
that the angular momentum predicted by the model that assumes a fully turbulent Hartmann layer ($R_T=0$) is significantly larger than the experimental 
values when $R<279$. This discrepancy between 
numerical and experimental values then diminishes rapidly as soon as 
$R\gtrsim279$. This reflects the behaviour of the mixing-length model for the
turbulent Hartmann layer: as the Hartmann layer becomes more and more turbulent,
 it becomes more and 
more accurate.\\
 Numerical simulations based on the model with $R_T=279$ become 
 very close indeed to those from the model based on a fully turbulent Hartmann 
layer in the limit of large $R$. Additionally, the model with $R_T=279$ 
performs a lot better than that with $R_T=0$ in  the limit of small $R$ where 
the Hartmann layers are laminar everywhere. In this last case, the model coincides with the SM82 model which slightly overestimated the angular momentum, compared to the experiment, as noted by \cite{delannoy99}.
When $R$ is of the order of 380 the model with threshold reproduces well 
the saturation observed in the experiment. Considering that the dissipation 
incurred by the metallic electrodes should imply that experimental values be a 
little lower than those returned by the model (as for large $R$), we must 
conclude that both models with $R_T=0$ and $R_T=279$ overestimate the 
dissipation by around 10\% in this transitional regime.\\
Finally, a handful of cases with $R_T=380$ were computed and they were found 
to differ very little from those at $R_T=279$, apart from a slightly better 
performance in the transitional regime. This is certainly an indication 
that the transitional regimes involve more complex mechanisms than a local 
threshold on the local friction.\\

\begin{figure}
\includegraphics[width=7cm]{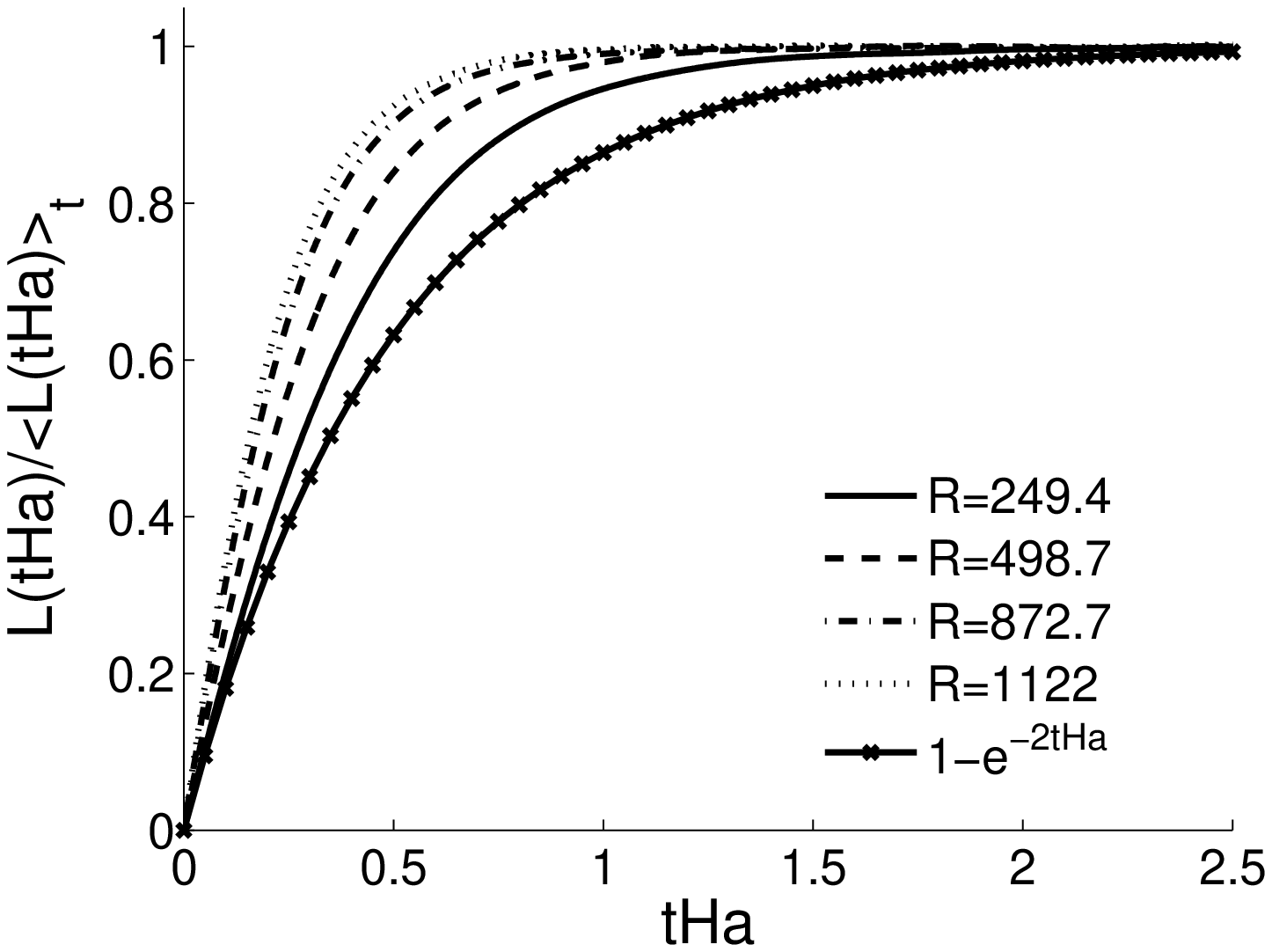}\\
\includegraphics[width=7cm]{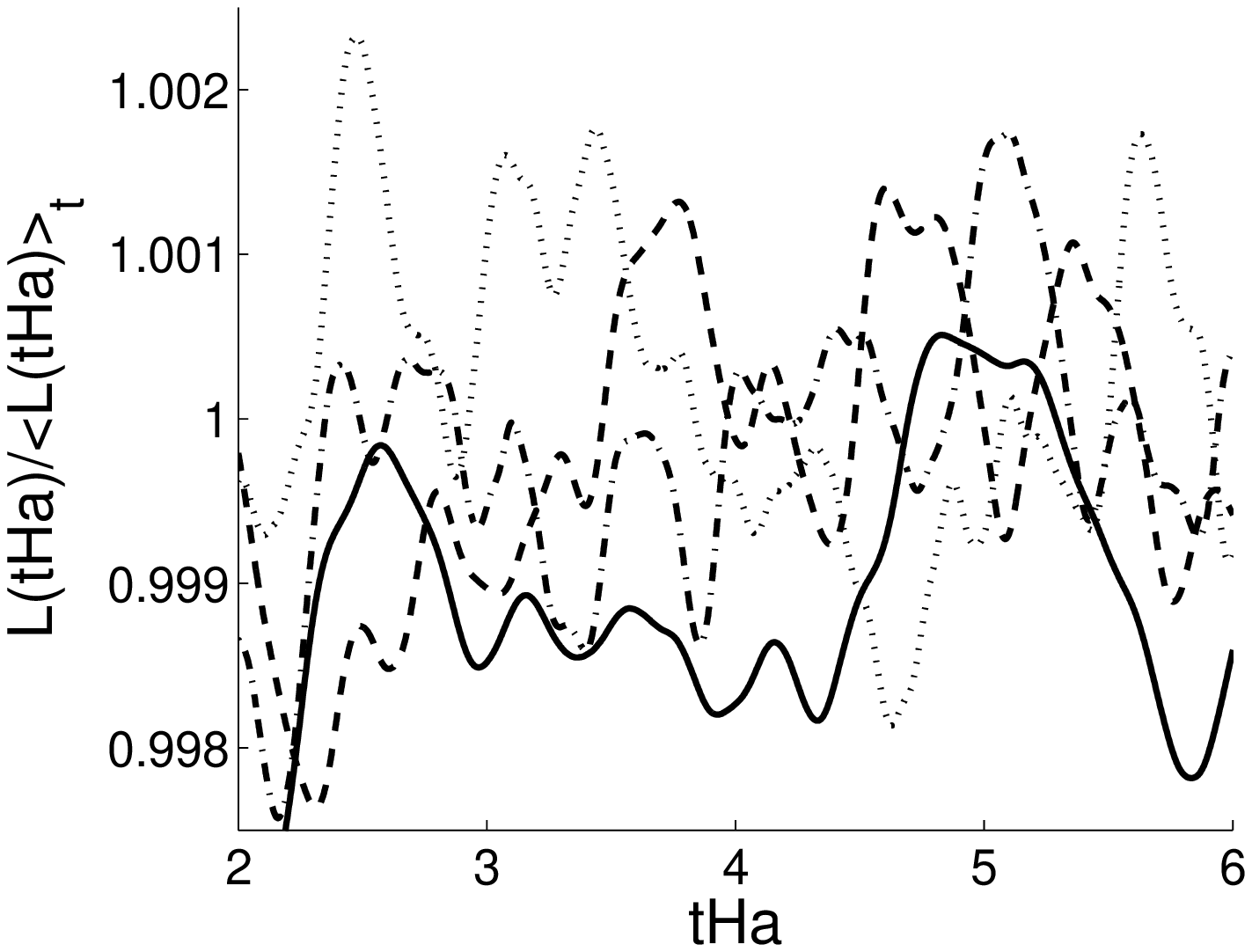}
\caption{Relative time-variations of the global Angular moment in MATUR under 
constant forcing, obtained from the model with $R_T=0$. Top: "spin-up" transient 
with the fluid initially at rest. The theoretical evolution of $L$ according to SM82 (axisymmetric) 
is represented to illustrate how turbulent friction shortens the flow reaction 
time. Bottom: fluctuations in the established regime.
$tHa$ is the non-dimensional time normalised by the Hartmann friction time, 
while $< \cdot >_t$ stands for time averaged quantities in the established 
state.}
\label{fig:l_uns}
\end{figure}

The time variations of the global angular momentum reveal a further two 
properties of the flow. Firstly, figure \ref{fig:l_uns} (top) shows that the 
transient time required to bring the flow from rest to an established state 
decreases with $R$, for $R_T=0$. This contrasts with quasi-2D flows with laminar Hartmann layers where the dimensional linear friction time $t_H$ is 
independent of the flow intensity. Secondly, the flow in the established regime 
exhibits erratic fluctuations of global angular momentum of a relative intensity
 that remains around $0.3\%$ through the range of parameters spanned here. 
Fluctuations of similar amplitude
 were  found in numerical simulations of MATUR performed with the SM82 model in 
cases where the Hartmann layers were laminar \cite{psm05}. Thus,  
although the turbulent Hartmann layer produces a lot more dissipation than its 
laminar counterparts, it doesn't eliminate the oscillations of the quasi-2D 
angular momentum, as the PSM model does.
\subsection{Radial profiles of azimuthal velocity}
\label{sec:velocities}
\begin{figure}

\includegraphics[width=6.25cm]{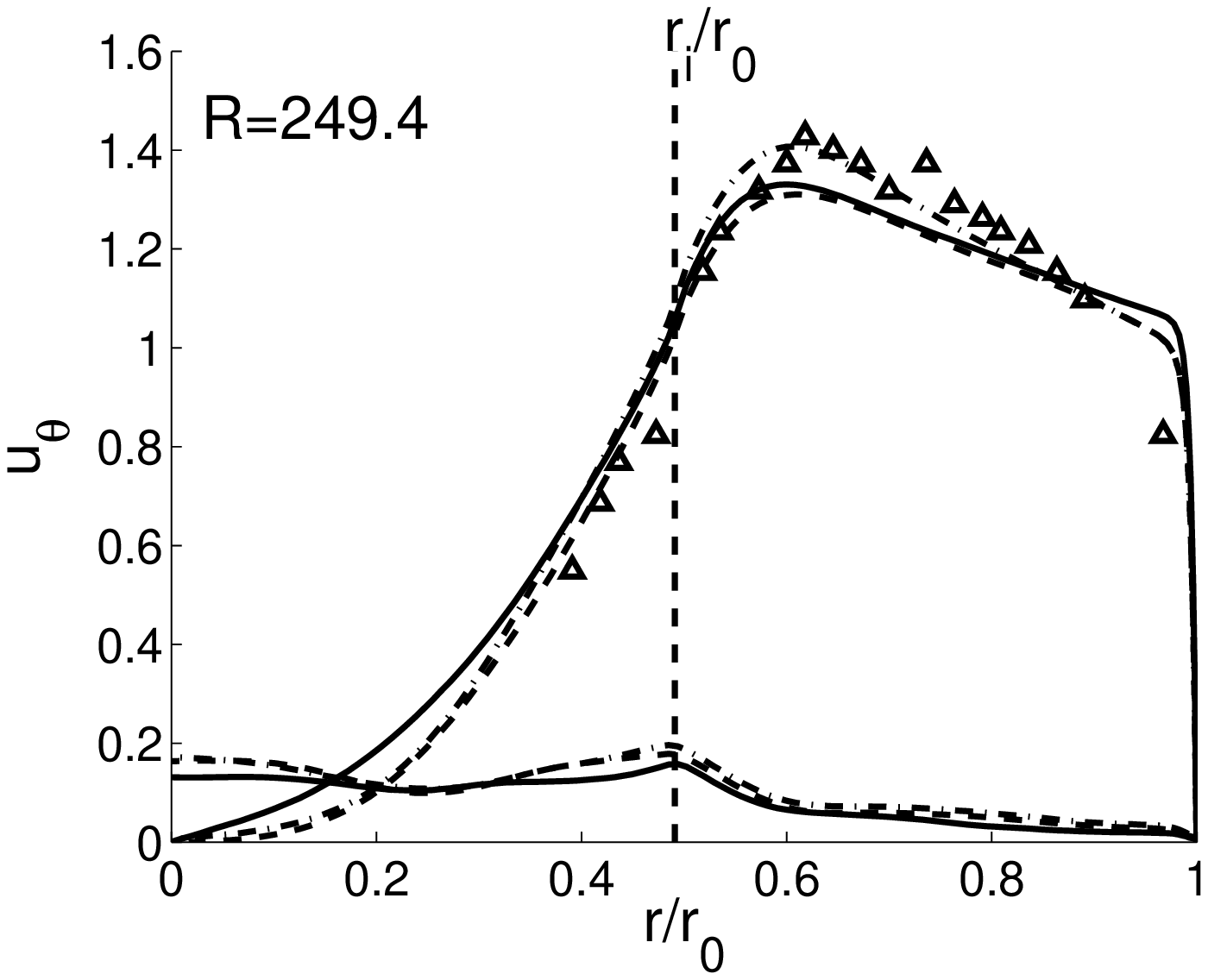}\\
\includegraphics[width=6.25cm]{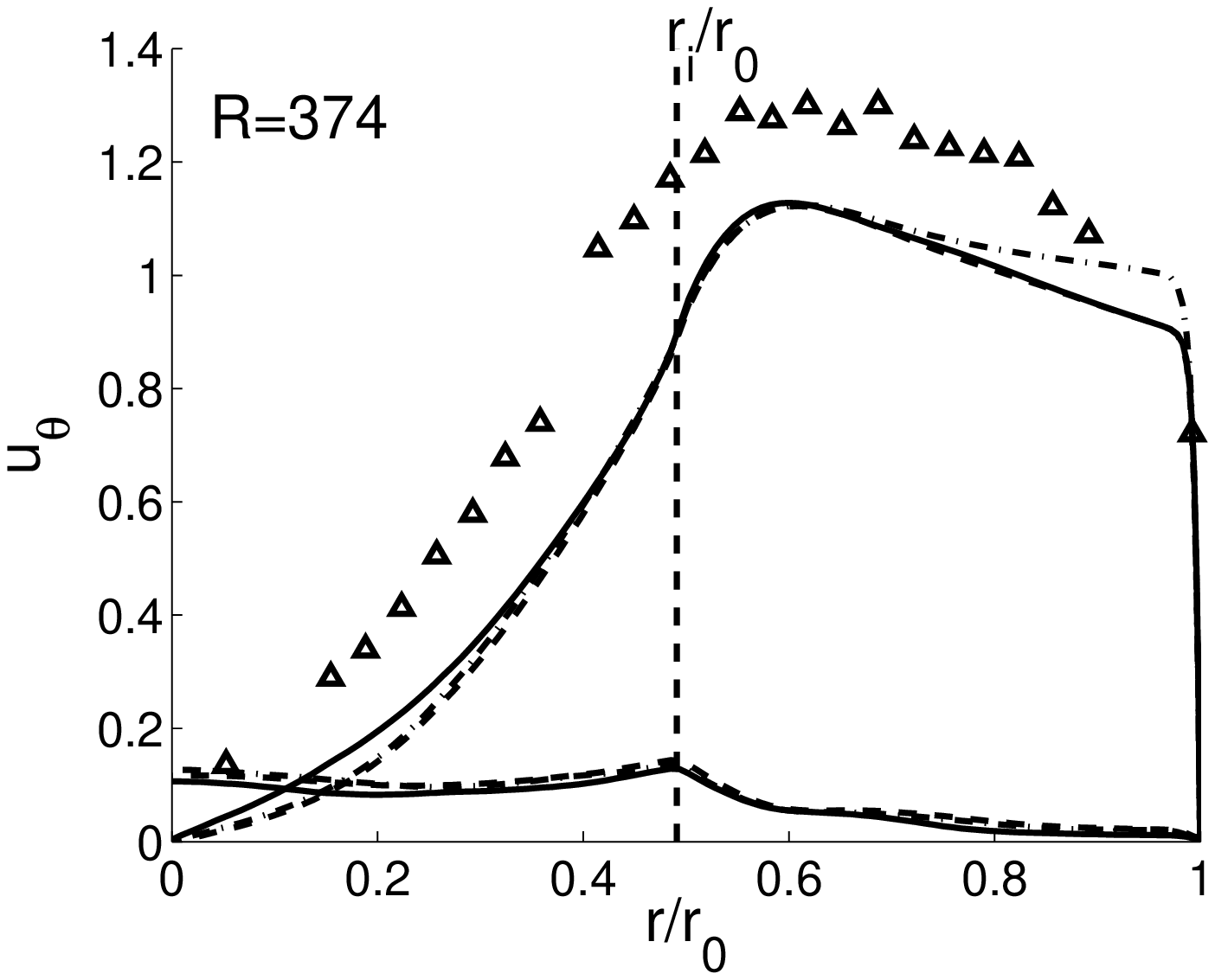}\\
\includegraphics[width=6.25cm]{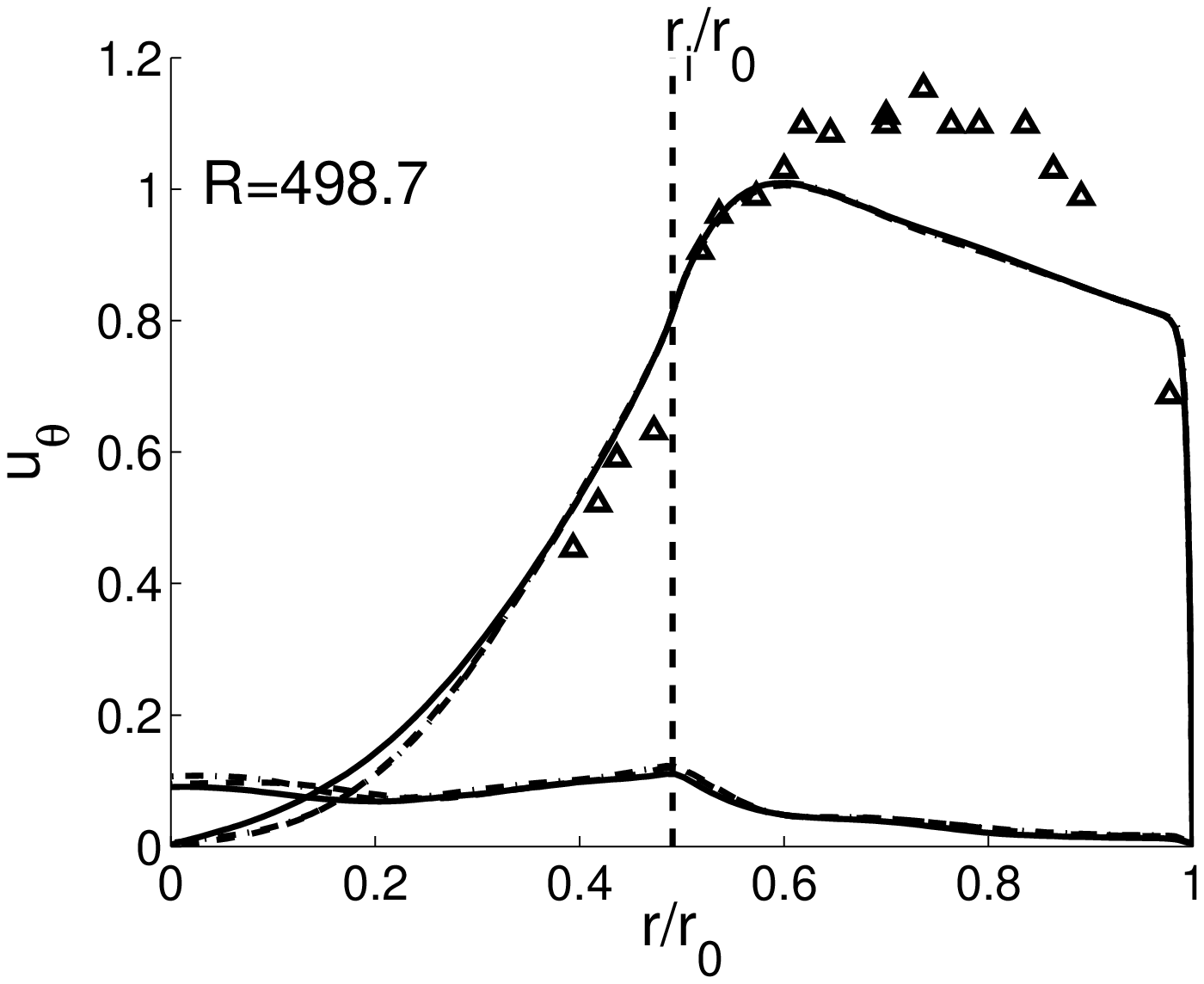}\\
\includegraphics[width=6.25cm]{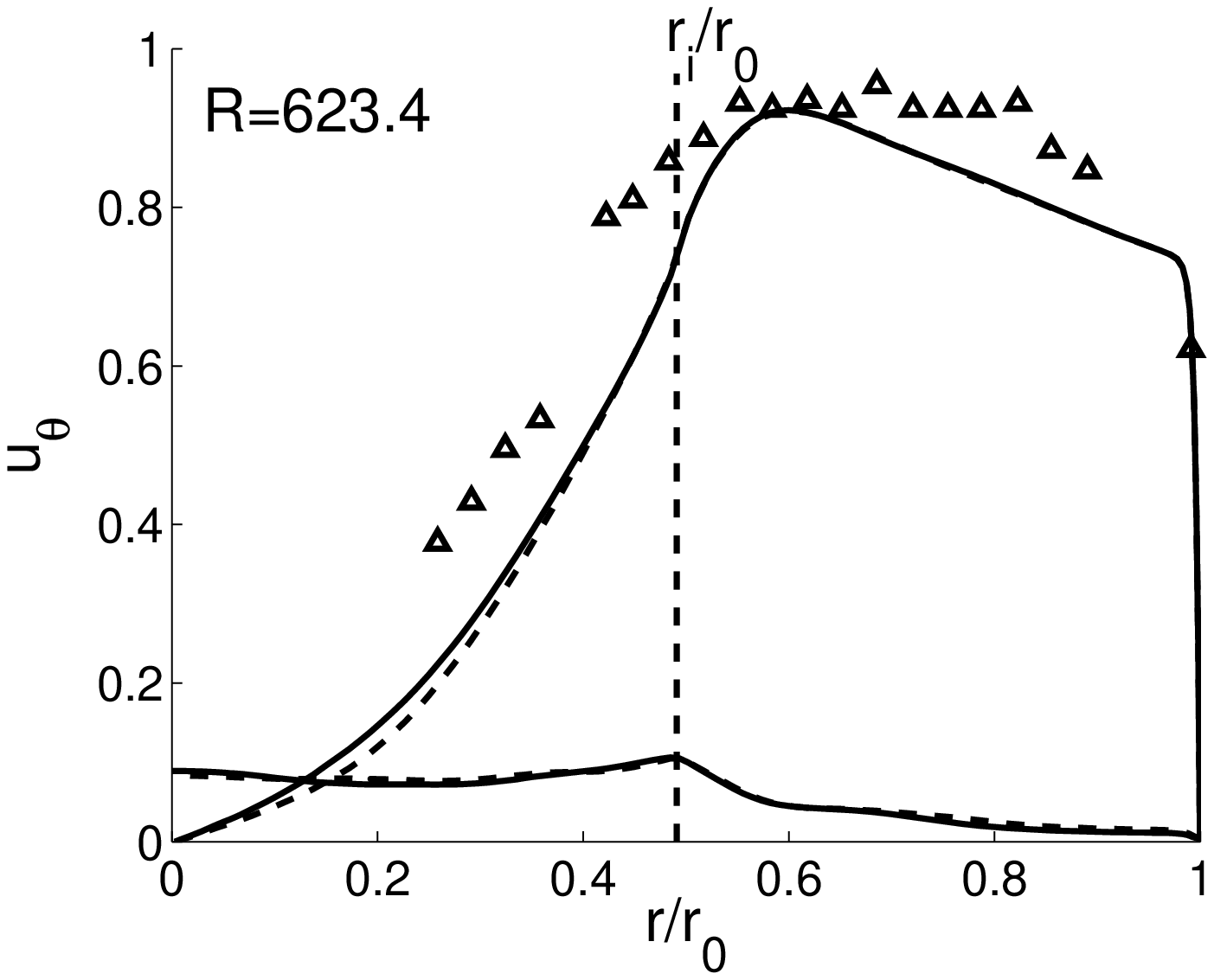}\\
\psfrag{Messadek}{\tiny $u_\theta$ from \cite{messadek01_phd}}
\psfrag{RT380}{\tiny 2D model $R_T=380$}
\psfrag{RT0}{\tiny 2D model $R_T=0$}
\psfrag{M2}{\tiny 2D model $R_T=0$, Mesh M$_2$}
\caption{Radial profiles of mean azimuthal velocity and RMS fluctuations of
azimuthal velocity (set of curves with values around 0.1). Legend is on figure 
\ref{fig:vtheta2}.}
\label{fig:vtheta}
\end{figure}

\begin{figure}
\includegraphics[width=6.25cm]{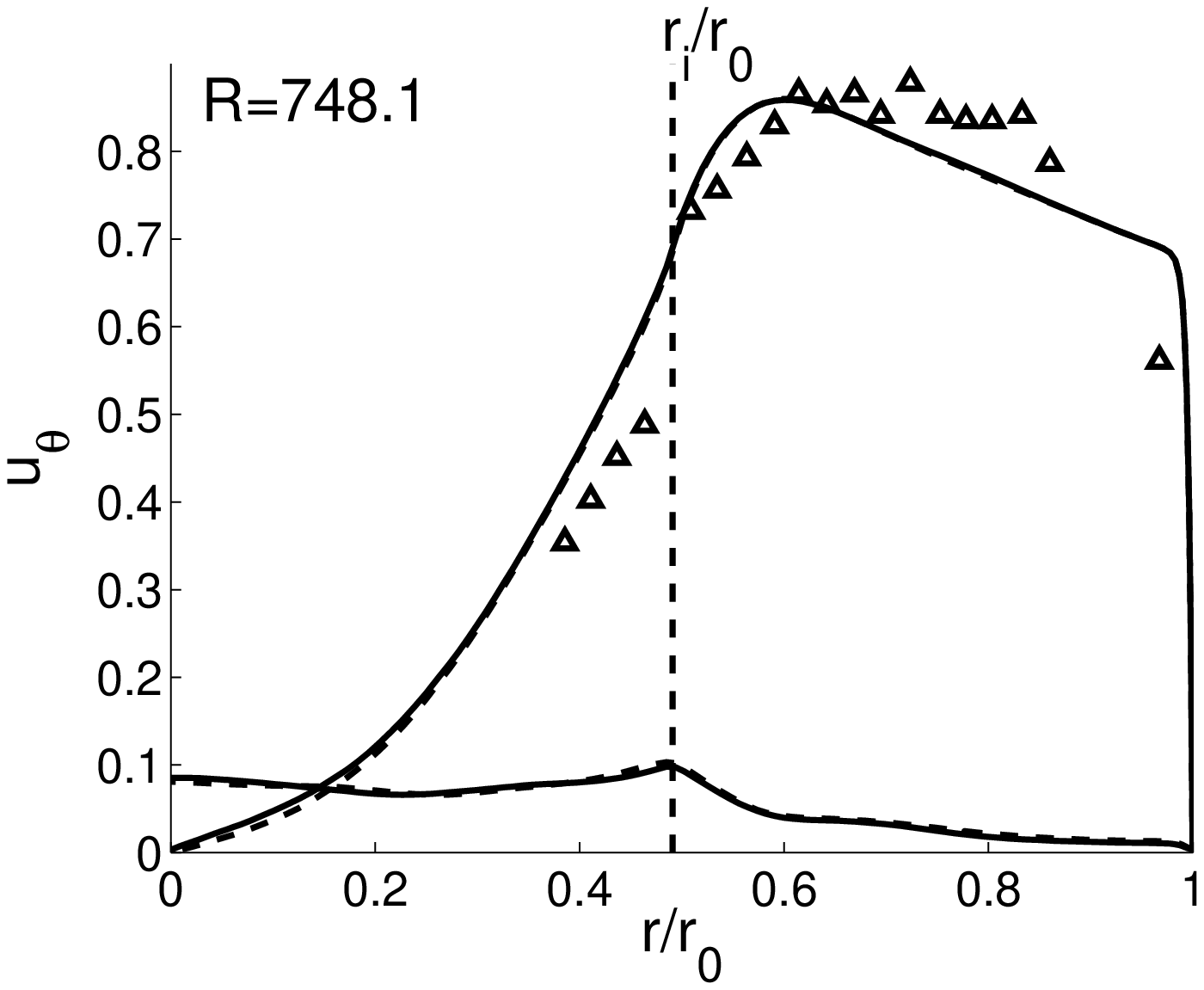}\\
\includegraphics[width=6.25cm]{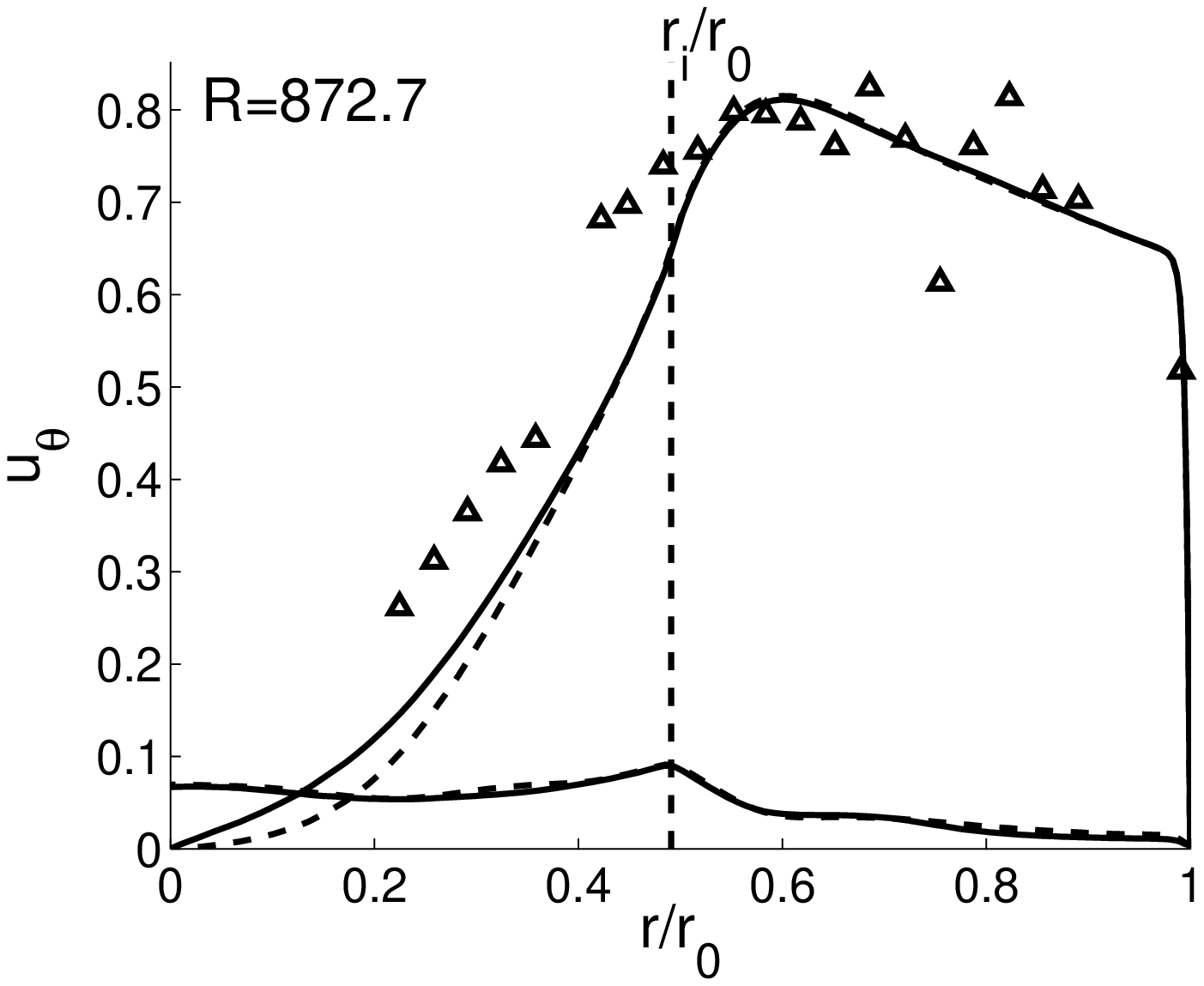}\\
\includegraphics[width=6.25cm]{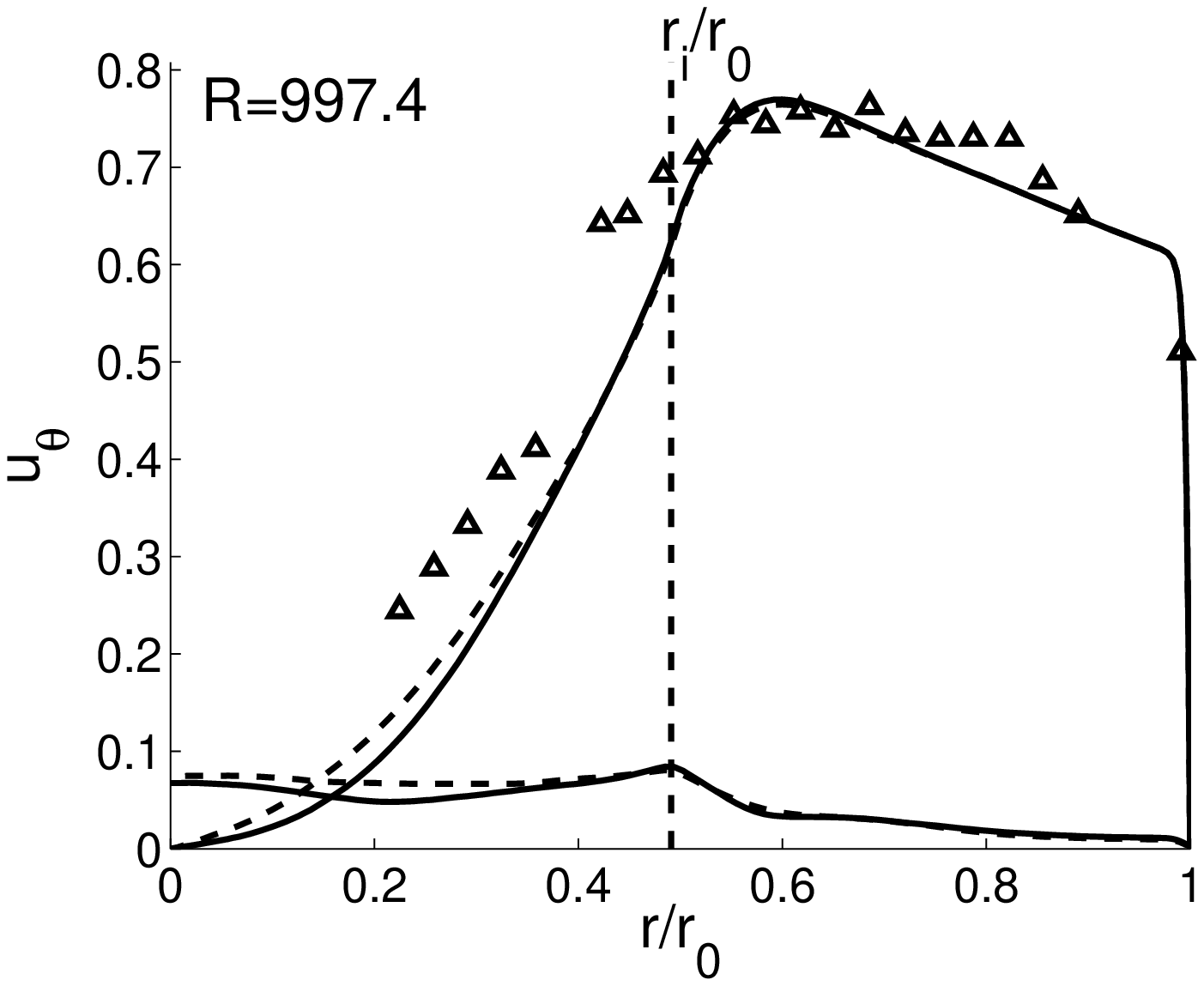}\\
\includegraphics[width=6.25cm]{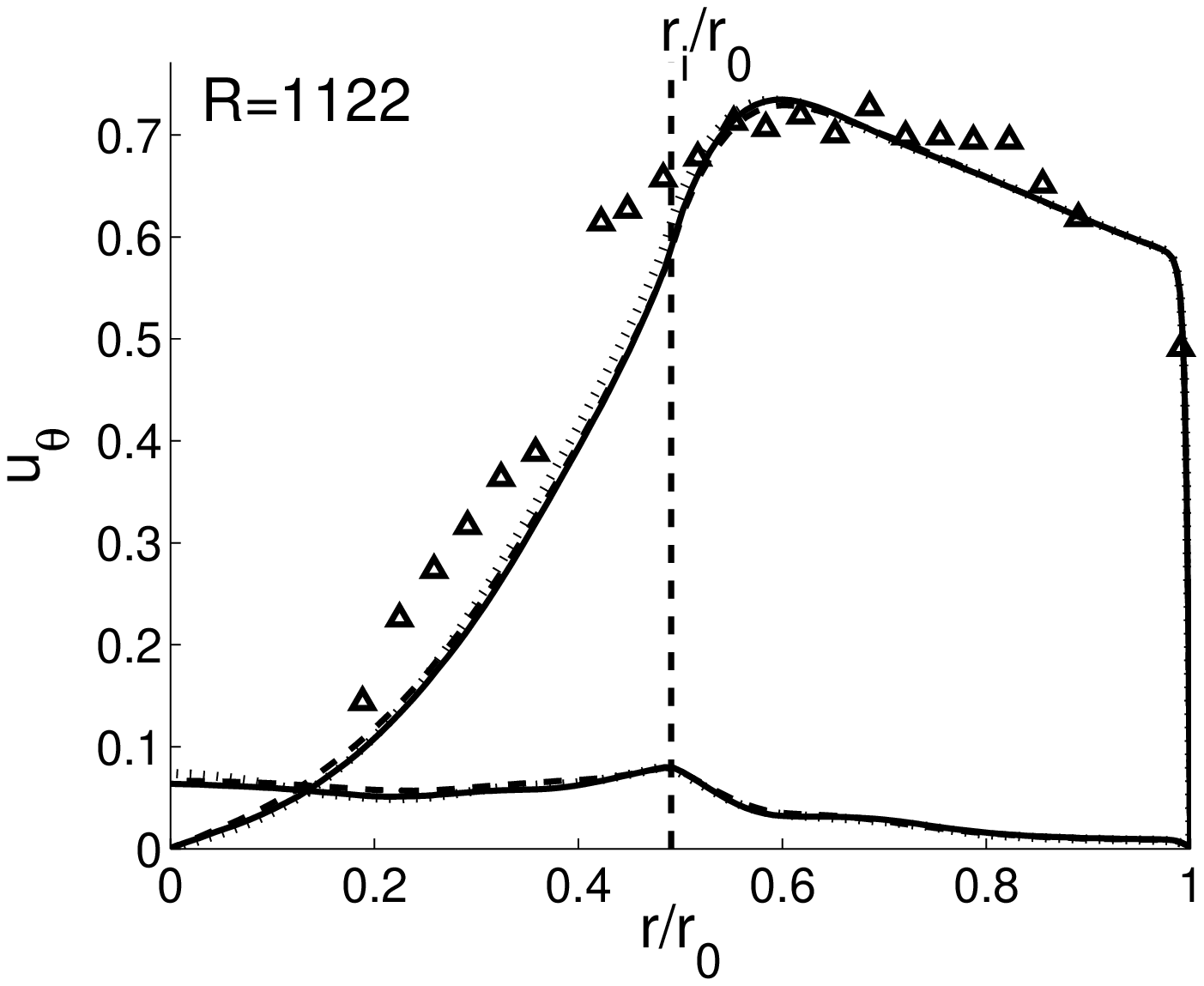}\\
\psfrag{Messadek}{\tiny $u_\theta$ from \cite{messadek01_phd}}
\psfrag{RT380}{\tiny 2D model $R_T=380$}
\psfrag{RT279}{\tiny 2D model $R_T=279$}
\psfrag{RT0}{\tiny 2D model $R_T=0$}
\psfrag{M2}{\tiny 2D model $R_T=0$, Mesh M$_2$}
\includegraphics[width=3cm]{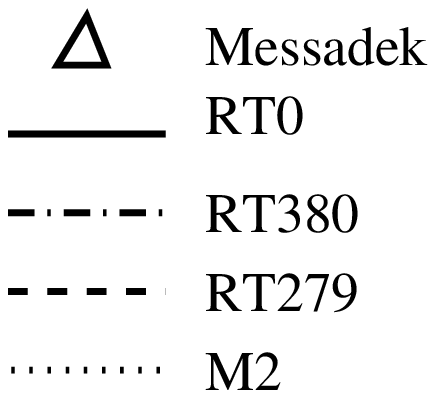}
\caption{Radial profiles of mean azimuthal velocity and RMS fluctuations of 
azimuthal velocity (set of curves with values around 0.1). Comparison between 
experimental, numerical results obtained with the models with threshold, 
without threshold, and on mesh M$_2$ (for $R=1122$ only).}
\label{fig:vtheta2}
\end{figure}

The radial profiles of time-averaged azimuthal velocity in  the established 
regime (figures \ref{fig:vtheta} and \ref{fig:vtheta2}) confirm the conclusions reached when analysing 
 the global angular momentum: the discrepancy between experimental and numerical
profiles decreases as $R$ increases and the Prandtl model becomes more accurate. For $R\gtrsim 700$, the error can hardly be distinguished from the 
experimental error. 
Even so, it seems that the turbulent model slightly underestimates azimuthal 
velocities in the central region for the larger values of $R$. 
 Furthermore, even in the most turbulent cases analysed here, the parameter 
$R(\Gamma/(2\tilde r_0))$ which is based on the linear estimate for the velocity $\Gamma/(2\tilde r_0)$, is of 1247. In this case, the actual maximum velocity in the flow is about half of $\Gamma/(2\tilde r_0)$, so a more realistic value of $R$ would be around 600, which is 
only mildly supercritical. Considering this, the performances of the 2D model 
are excellent. Furthermore, it can be noticed that there are only few 
experimental points in the vicinity of the wall at $r=r_0$. Since this region 
brings the highest contribution to the global angular momentum, the experimental error there might also be in part responsible for the residual difference in 
angular momentum at high $R$ between our model and the experiment. In spite of 
this minor uncertainty, the fact that both global and local quantities measured in MATUR are closely recovered over a wide range of parameters by the numerical simulations of our model is certainly a good evidence that the extra dissipation observed at $Ha=132$ and $Ha=212$ is indeed due to the turbulent state of the
Hartmann layers.\\
The models with $R_T=279$ and $R_T=380$ improve on that with $R_T=0$ in that 
they very accurately render regimes where the Hartmann layer is laminar (case 
$R=249.4$). In the transitional regime around $R=380$, even though both models 
are able to reproduce the curve $L(R)$, they underestimate the 
actual velocity of the flow by up to $15\%$ in the outer region 
$r_i\leq r\leq r_0$ (see figure \ref{fig:vtheta} for $R=374$ and $R=498.7$, 
for which the discrepancy between models and experiment is most conspicuous). 
For $R=374$, the model with $R_T=380$ yields higher velocities in the vicinity 
of the outer wall at ($r=r_0$) 
than that with $R_T=279$, because the value of $\bar u$ such that 
$R(\bar u)=380$ is reached between $r_i$ and $r_0$.  When only a small 
part of the flow is subject to turbulent friction (case with $R=249.4$), the 
model with $R_T=380$ performs better than that with $R_T=279$. This is an 
evidence that the Hartmann layer is 
almost entirely laminar in the experiment in this regime. Since with a 
threshold of $R_T=380$, only a very narrow region around $r/r_0\simeq0.6$ 
experiences turbulent friction, this value of $R_T$ turns out to yield more 
realistic results than $R_T=279$.  As expected, for higher values of $R$, the 
profiles of  
velocity and velocity fluctuations obtained with $R_T=279$ and $R_T=380$ with 
threshold depart little from the model with $R_T=0$. The profiles obtained 
from both models with threshold differ even less from each other, 
to the point where they can't be distinguished on the graph. Overall, the 
model with $R_T=380$ can be deemed valid whenever $R\lesssim 300$ or 
$R\gtrsim600$.\\
\subsection{Velocity fluctuations}
\begin{figure}
\psfrag{v}{$u$}
\includegraphics[width=8.25cm]{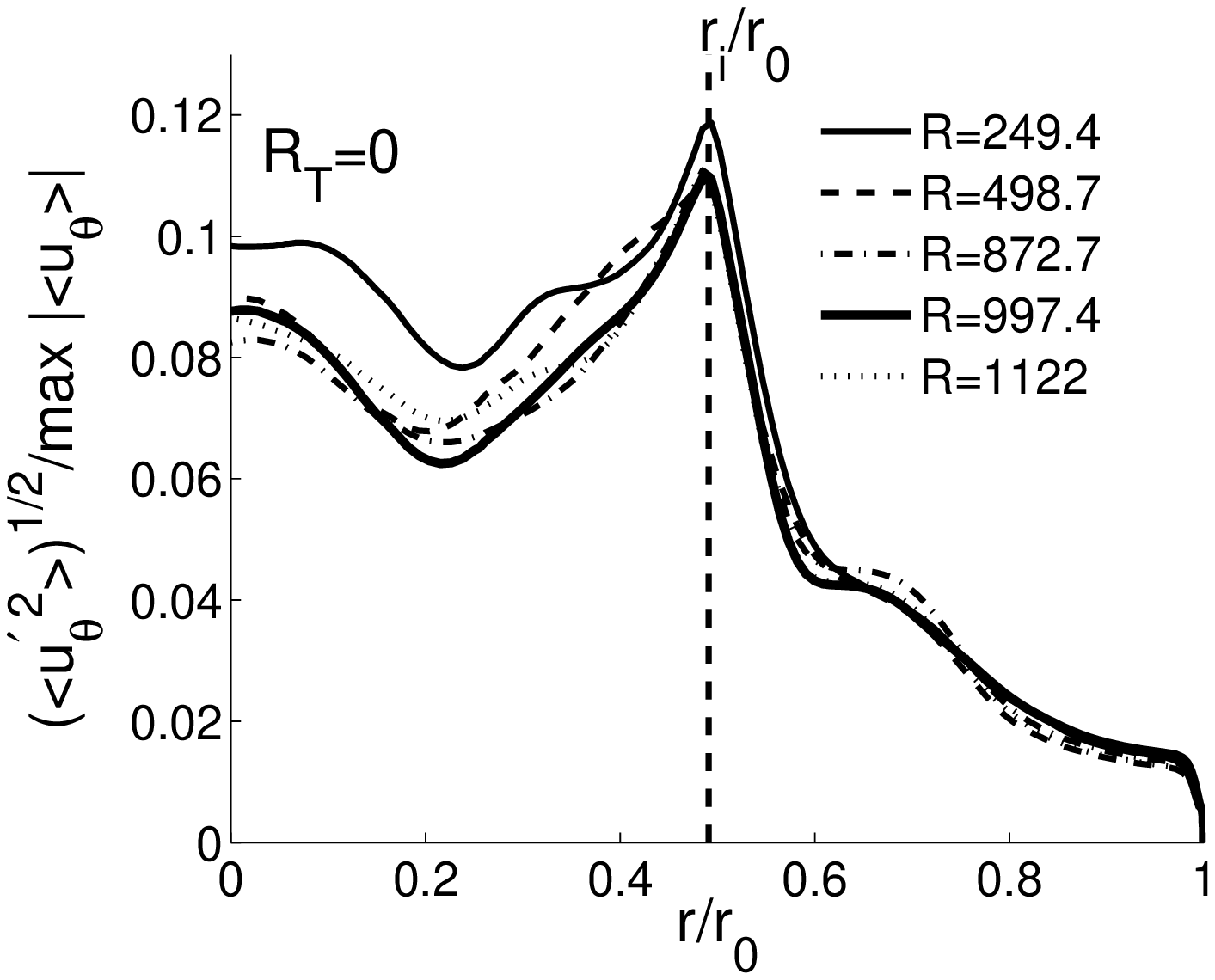}
\includegraphics[width=8.25cm]{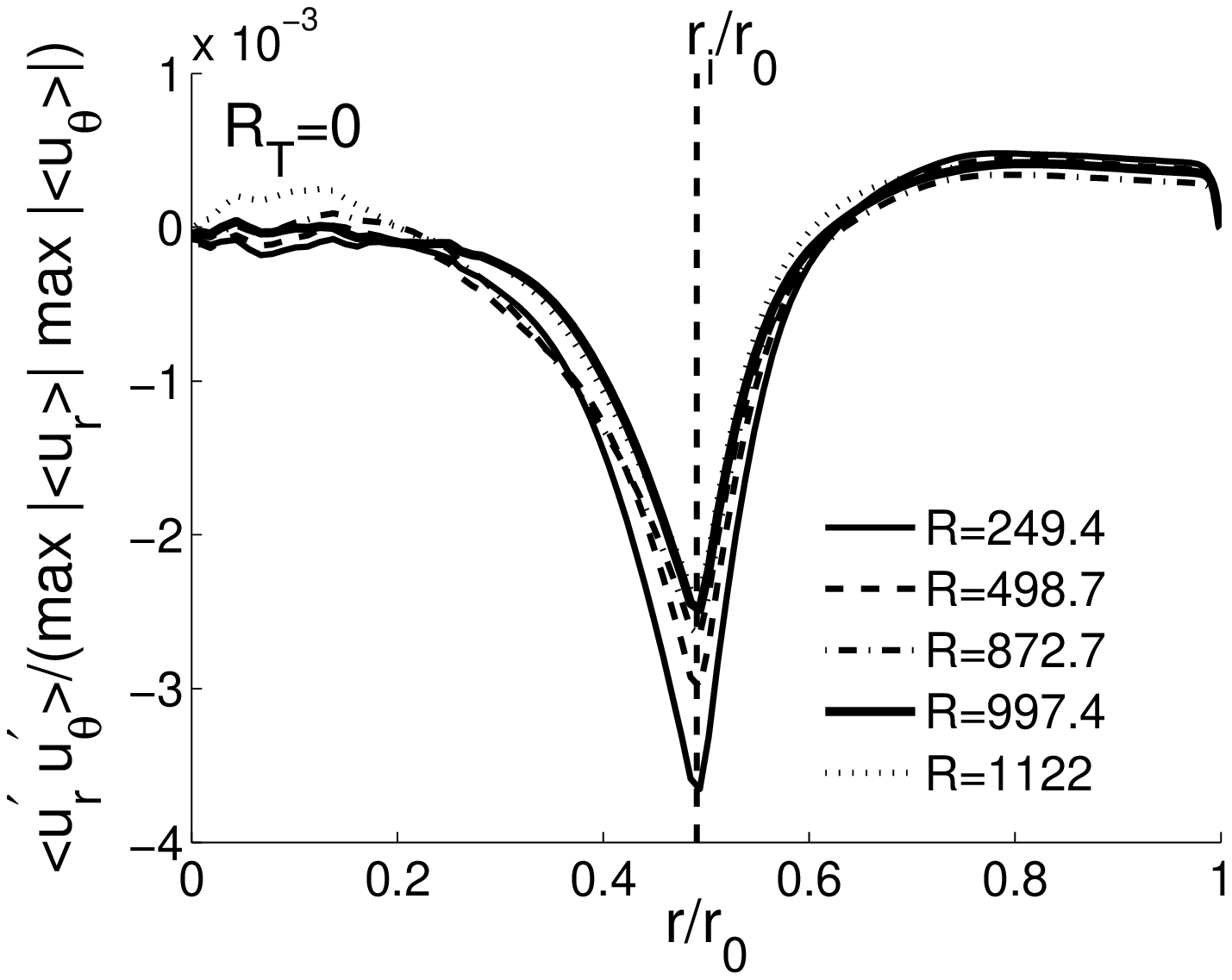}
\caption{RMS of relative azimuthal velocity fluctuations (top) and correlations 
of azimuthal and radial velocity fluctuations, normalised by maximum average 
velocities (bottom). Both graphs were obtained from simulations of the 2D model 
 with $R_T=0$.}
\label{fig:fluctcorr}
\end{figure}

The RMS averages of absolute azimuthal velocity fluctuations are reported on 
figures \ref{fig:vtheta} and  \ref{fig:vtheta2} and their relative counterpart are gathered on figure 
\ref{fig:fluctcorr} (top). All curves exhibit a more or less triangularly 
shaped maximum at the location of the current injection electrodes. This 
reflects the 
passage of the large structures that result from the merging of small vortices 
generated by the instability of the circular free shear layer at $r=r_i$. The 
width of the triangle gives an idea of the size of these structures. In all 
cases, fluctuations are significantly higher in the region $r<r_i$ than for 
$r>r_i$. This indicates that, as seen from the contours of vorticity on figure 
\ref{fig:w_contours}, once released from their region of origin, these large 
structures drift towards the centre of the domain rather than towards the 
external wall, unlike in cases where $r_i/r_0=0.84$ \cite{psm05}. The shape of the 
profile remains the same when $R$ increases, while the relative intensity of 
the fluctuations decreases only slightly. We shall see thereafter that this 
behaviour mostly results from 
the competition between a more intense flow, which drives more intense relative 
velocity fluctuations and the turbulent Hartmann friction, which damps them.
Indeed, the latter increases several times more than its laminar counterpart 
with the flow intensity.\\
The radial profiles of the relative correlation between radial and azimuthal 
velocity fluctuations shown on figure \ref{fig:fluctcorr} (bottom) give a good 
measure of the turbulent intensity. The general aspect of these curves presents 
some interesting features: 
for $r<r_i$, where $<u_\theta^{\prime 2}>^{1/2}$ keeps relatively high values, 
the correlations drops to zero. Conversely, in the 
region $r>r_i$, where $<u_\theta^{\prime 2}>^{1/2}$ dropped, the correlations 
exhibit a moderately high, positive value. This reflects the difference in 
the nature of turbulence between these two regions already noted in section 
\ref{sec:aspect}: for $r<r_i$, 
fluctuations are fed by large structures drifting to the centre.
By contrast, fluctuations in the outer region ($r>r_i$) are the trace of 
azimuthal streaks 
of vorticity that originate from the tail of the large structures. These are 
stretched by the shear and transported outwards.\\
Furthermore, the correlations of relative radial and azimuthal velocity fluctuations 
decrease more noticeably with $R$ than the RMS velocity fluctuations (this is 
partly due to the former being a quadratic function of the velocity, while the latter  are linear). 
They then stabilise at nearly the same value for $R\gtrsim997$. This diminution 
of turbulence intensity reflects that the turbulent Hartmann layer friction, 
which increases non-linearly with $R$, absorbs an ever increasing fraction 
of the energy injected in the flow at the expense of quasi-2D turbulent 
fluctuations. \\
This nonlinear variation of $\tau_W$ with $R$ also explains that the region 
characterised  by negative 
correlations or by the triangular-shaped maximum of RMS velocity fluctuations 
doesn't appreciably increase in size with $R$. If anything, it even slightly 
narrows. Since it is essentially determined by passing large structures, this 
phenomenon can be understood by noticing that the size 
of these structures is limited by boundary layer friction: if $U_L$ is the 
typical self rotation velocity of a vortex of size $L^{\rm turbulent}$  
(\emph{resp.} $L^{\rm laminar}$) when the Hartmann layers are turbulent 
(\emph{resp.} laminar), then any vortex with a turnover time 
$L^{\rm turbulent}/U_L$ 
(\emph{resp.} $L^{\rm laminar}/U_L$) higher 
than the typical Hartmann layer friction time is dissipated 
\cite{chertkov07_prl}. This determines their scaling as:
\begin{eqnarray}
\frac{L}{H}^{\rm turbulent}&\sim&\frac{R}{f(R)}\left(\frac{U_L}{U}\right)^{\rm turbulent}
\label{eq:large_scales} \\
&\leq&\frac{R}{f(R)}\left(\frac{U_L}{U}\right)^{\rm laminar}
=\frac1{f(R)}\frac{L}{H}^{\rm laminar} \nonumber\\
&<&\frac{L}{H}^{\rm laminar}. \nonumber 
\end{eqnarray}
For turbulent Hartmann layers, $f(R)$ is greater than unity and increases 
monotonically (see figure \ref{fig:tau_nondim}). Furthermore, since ${U_L}/{U}$ 
is roughly the 
intensity of the azimuthal velocity fluctuations, it decreases a little with 
$R$ and it is smaller when the Hartmann layers are turbulent than when they 
are laminar. The scaling (\ref{eq:large_scales}) thus shows that the 
increasing turbulent friction opposes the increase in size of the large scales 
with $R$ and that those are therefore smaller than when the Hartmann layers 
are laminar. This explains why the region where $<u'_ru'_\theta>$ is negative 
doesn't widen with $R$. It also explains that the thickness of the free shear 
layer at $r=r_i$,  which the large structures conveyed by the flow also 
determine, remains seemingly unchanged as $R$ increases (This can be seen on 
figures \ref{fig:vtheta} and \ref{fig:vtheta2}). By contrast, 
\cite{messadek02_jfm} found that when Hartmann layers were laminar, and 
boundary layer friction was less intense, the thickness of this layer slowly 
increased as $R^{1/2.2}$.\\
Finally, it should be noted that both types of fluctuations obtained with the 
model at $R_T=279$ exhibit essentially the same behaviour as those from the
model at $R_T=0$. 
%
%
%
%
\section{Conclusion}
We have established a 2D model that applies to channel flows under transverse 
magnetic fields with turbulent Hartmann layers. Numerical simulations of the MATUR experiment based on it gave strong evidence that the previously 
unexplained level of dissipation observed at $Ha=132$ and $Ha=212$ was 
caused by turbulence in the Hartmann layers.\\
Unlike its predecessors, which 
account for laminar Hartmann layers, the new model is not rigorously derived 
from first principles but relies instead on the equations for the Hartmann 
layer friction based on Prandtl's assumption proposed by \cite{albouss00}. 
Nevertheless, as soon as the Reynolds number based on the Hartmann layer 
thickness exceeds about 600, 2D numerical simulations of this model reproduce 
the experimental results from \cite{messadek02_jfm} with discrepancies below 
10\% on the global angular momentum and an error on local velocities that falls within the experimental error. The parametric analysis for $124<R<1247$ 
performed in this work reveals that the precision of the model increases with 
$R$, a feature it inherits from \cite{albouss00}'s and Prandtl's models. This sheds an 
even better light on the precision of the results obtained here, since in terms 
of the velocity actually achieved in the flow, the highest value of $R$ reached 
here barely exceeded 600, which according to the work of \cite{moresco04} is 
only mildly supercritical, in terms of the transition to turbulence in the 
Hartmann layer.\\
We have also introduced an admittedly artificial variant of our model where the 
boundary layer friction reverted to its laminar value below a threshold value 
of $R_T$. $R_T$ was tuned either to 279, value at which laminar and turbulent 
frictions coincide, or to the value of 380 found by \cite{moresco04} and 
\cite{krasnov04} for the transition to turbulence in the Hartmann layers in a 
rectilinear channel flow. Although the models with thresholds cannot precisely 
render the transitional regimes $300\lesssim R\lesssim600$ where neither of 
the 2D models based on fully laminar or fully turbulent Hartmann 
layers are meant to operate, they gather these two models in a single one.
The results obtained with either thresholds don't differ a great deal, although
only the model with $R_T=380$ recovers well the experimental values of the 
global angular momentum, even in transitional regimes of the Hartmann layers 
 ($R\simeq 380$).  Threshold models therefore constitute a useful extension 
of the fully turbulent model, particularly for flows where the state of the 
Hartmann layers may not be known \emph{a priori}.\\ 
Despite not sharing the asymptotic pedigree of their predecessors (SM82 and PSM), the new class of shallow water models we introduced  not only offers the 
same flexibility and simplicity but also the same level of performance. In 
this regard, it makes it now possible to simulate flows as complex as those in 
MATUR, where both three-dimensional boundary layer turbulence and quasi-2D 
turbulence coexist a low computational cost. This was previously not possible 
with either SM82 or PSM since these models are restricted to flows where 
Hartmann layers are laminar. These new models now make extensive 
parametric analyses of a wide new class of flows with turbulent Hartmann layers 
easily accessible. Such an analysis would indeed incur very large computational costs if carried out with 3D simulations where the Hartmann layer would be meshed.\\
It is precisely such a parametric analysis that has allowed us to reveal two
important properties of quasi-2D flows with turbulent Hartmann layers: firstly,
turbulent friction restricts the size of the large scales, compared to its 
laminar counterpart. Secondly, it has a stabilising effect on the quasi-2D 
flow, as it dissipates an increasingly high fraction of the 2D turbulent energy 
when the flow is driven more intensely.\\

The authors are grateful to Thierry Alboussi\`ere and Pablo Moresco for their 
fruitful input during the conduct of this work. They are also grateful to 
referee 2, whose remarks have greatly helped to improve the precision of the 
models. 


\end{document}